\documentclass[useAMS,usenatbib]{mn2e}

\usepackage{graphicx}

\usepackage[dvips]{color}

\title[The metal abundance distribution of the oldest stellar component in Sculptor]
{The metal abundance distribution of the oldest stellar component  
in the Sculptor dwarf 
spheroidal galaxy\thanks{Based on data 
collected at the
European Southern Observatory, proposal number 71.B-0621}}

\author[Clementini et al.]
{G. Clementini$^{1}$\thanks{E-mail: gisella.clementini@bo.astro.it;   
ripepi@na.astro.it; angela.bragaglia@bo.astro.it; 
fiorenzano@pd.astro.it; held@pd.astro.it; gratton@pd.astro.it},
V. Ripepi$^{2}$,
A. Bragaglia$^{1}$,
A.F. Martinez Fiorenzano$^{3,4}$,\newauthor
E.V. Held$^{3}$, and R.G. Gratton$^{3}$\\
$^{1}$INAF - Osservatorio Astronomico di Bologna, via Ranzani 1, I-40127
Bologna, Italy\\
$^{2}$INAF - Osservatorio Astronomico di Capodimonte, via Moiarello 16, 
I-80131 Napoli, Italy\\
$^{3}$INAF - Osservatorio Astronomico di Padova, vicolo dell'Osservatorio 5, I-35122 
Padova, Italy\\
$^{4}$Dipartimento di Astronomia, Universit\`a di Padova, vicolo dell'Osservatorio 2, 
I-35122 Padova, Italy\\
}

\begin{document}

\date{Received 2005 ; in original form 2005}

\pagerange{\pageref{firstpage}--\pageref{lastpage}} \pubyear{2005}

\maketitle

\label{firstpage}
\label{lastpage}

\begin{abstract}

Low resolution spectroscopy 
obtained with FORS2 at the Very Large Telescope (VLT) has been used to measure 
individual metal abundances ([Fe/H])
 for 110 variable stars, including 107 
RR Lyrae stars and 1 Anomalous Cepheid, and 
trace the metal distribution of the oldest stellar component in the 
Sculptor dwarf spheroidal galaxy. The 
RR Lyrae stars 
are spread over a 15$\arcmin \times 15 \arcmin $ area around the galaxy centre.  
Their metallicities have an average value of [Fe/H]=$-1.83 \pm 0.03$\ 
(r.m.s. 0.26 dex)   
and cover the metallicity range $-2.40<$[Fe/H]$<-0.85$  
(on the scale of \citealt{zw84}), 
but there is only one 
variable having [Fe/H]$>-1.3$.
The star-to-star
scatter is 
larger than   
typical errors on individual metallicities ($\pm 0.15-0.16$~dex),
indicating a real
spread in metal abundances.
The radial velocities measured from the RR Lyrae spectra have a dispersion
of 12.9 km s$^{-1}$. This value is consistent with the dispersion 
derived by \citet{tol04} for metal-poor red giants associated to the blue horizontal branch 
stars in Sculptor. Along with the metallicity
distribution these results suggest that most of the RR Lyrae stars in Sculptor
arise from the same burst of stellar formation that produced the metal-poor component 
giving origin to the galaxy blue horizontal branch.
The metal-rich red horizontal branch population found to be centrally 
concentrated 
only produced a few (if any) of the RR Lyrae stars in our sample.
The spectroscopic metallicities were used along with the apparent luminosities 
to
study the luminosity-metallicity relation
followed by the RR Lyrae stars in Sculptor, for which we derive a 
shallow slope of 0.09 mag/dex. This result can be due to a high
level of evolution off the zero age horizontal branch of the RR Lyrae
stars in this galaxy, again in agreement with their origin from the
blue horizontal branch population.

\end{abstract}


%

\section{Introduction}
In hierarchical merging scenarios dwarf galaxies are thought to be the bricks 
from which larger
galaxies were assembled. The study of the star formation history (SFH) and 
of the chemical evolution  
of presently existing dwarf galaxies is then extremely important for a 
proper understanding 
of the formation and evolution of the larger systems that this type of galaxies 
may have contributed  
to build in the past. Colour magnitude diagrams (CMDs) reaching the faint main 
sequence 
turn-off (TO) of the oldest stellar components 
are the most traditional and reliable way to derive the SFH of any 
individual dwarf galaxy. However, these 
detailed studies, possible so far mainly for the
dwarf members of the Local Group (LG), require very time consuming 
observations.

The RR Lyrae variables,
being about 3 magnitudes brighter than coeval TO stars ($t> 10$ Gyrs),
are much easier to observe. 
They  offer an excellent tool for tracing
the oldest stellar populations,  
and therefore the epoch of galaxy formation, in composite systems such as the
resolved LG dwarf galaxies.
Recent work by several groups has 
lead to the discovery of RR Lyrae stars 
in increasing numbers of LG dwarf 
galaxies (e.g. Leo~I and II,
IC~1613, Fornax, And~VI, NGC6822:  
\citealt{h01,sm00,do01,bw02,pritzl02,clem03a},  
just to 
mention a few of them). 
An early stellar
population, nearly coeval to the old Galactic globular clusters, has been
found in the majority of LG galaxies, irrespective 
of their star
formation histories.  This indicates that all LG dwarfs started
forming stars at an early epoch, $\sim$13 Gyr ago (e.g. Held et al. 2000). 

The Sculptor dwarf spheroidal galaxy is no exception to this general
trend. 226 RR Lyrae stars and 3 Anomalous
Cepheids have been detected in a 15$\arcmin \times 15 \arcmin $ area
around the galaxy centre
 by \citet{kal95}
who published multi-epoch photometry for all of them.
\citet{kal95} also found that the period distribution of the RRab stars shows a sharp cut-off at $P=0.475$ d implying a
metallicity of [Fe/H]$\leq -1.7$ (on the Zinn \& West 1984 scale),
and that the dispersion
of the average $V$\ magnitudes 
is most likely due to the metallicity spread exhibited
by the stars in this galaxy. 
Similarly, \citet{kov01} found $\langle$[Fe/H]$\rangle\sim
-1.5$ with a large dispersion from about $-2.0$ to $-0.8$ dex 
(on the metallicity scale by \citealt{jurcsik95}),
from the Fourier
decomposition of the light curves of the RRab stars. 

Indeed, Sculptor dwarf spheroidal (dSph) has long been known,
from photometric studies, to have a large 
metallicity
spread 
and bimodality in the metallicity and 
spatial distribution of its horizontal branch (HB) stars \citep{maj99, hurley99, 
harbeck01, 
rizzisait75, babusiaux05}.
Very few spectroscopic determinations
of the metal abundance of Sculptor stars existed so far
(\citealt{nb78,tol01}, 2003).
\citet{gei05}, from high resolution spectroscopy of  red giants,
find 
[Fe/H] values in the range from $-2.10$ to $-0.97$ dex, confirming
a large metallicity spread in Sculptor. 
Two distinct ancient populations showing an abrupt
 change in the [Fe/H] distribution at about 12 arcmin from the galaxy
 center are found in Sculptor by \citet{tol04} based on FLAMES@VLT low resolution
spectroscopy and WFI imaging of the galaxy.
The two components have also different spatial distribution and velocity
dispersion.

In this paper we present metal abundance determinations based on multi-slit low resolution spectroscopy 
obtained with FORS2 at the VLT for  
more than a hundred RR Lyrae stars in Sculptor. 
The variables are spread over a 
15$\arcmin \times 15 \arcmin $ area around the galaxy centre, thus 
being almost coincident with the internal region of Sculptor 
where \citet{tol04} find segregation of red HB stars.
The knowledge of the {\it metal
distribution} of the RR Lyrae population allows to put important
constraints on the early star formation and chemical evolution histories of
the host galaxy, by removing  
the age-metallicity degeneracy:  
the earliest measurable data
point for the chemical enrichment history of Sculptor 
can be determined with accuracy. 

\begin{figure*} 
\includegraphics[width=18cm]{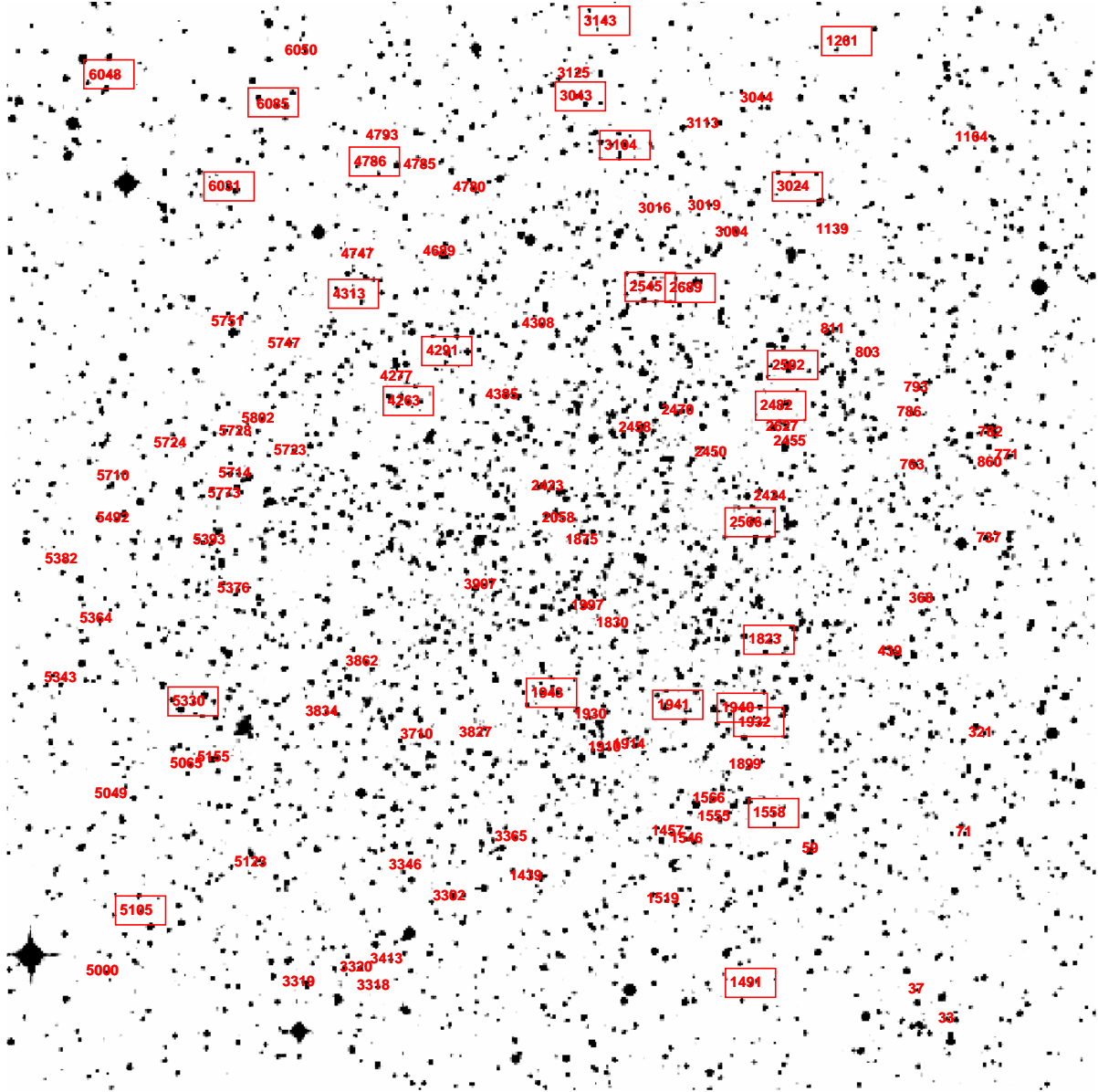}
\caption[]{Position of the observed variables on a
$16\arcmin \times 16 \arcmin$ map of the centre of 
Sculptor dSph galaxy.
North is up and East is to the left. Variables 
are identified according to \citet{kal95} identifiers. Also shown
(in blue in the electronic edition of the journal) are the four red giants 
recently analyzed by
\citet{gei05}. Boxes mark the RR Lyrae stars for which we measured
metallicites [Fe/H]$> -$1.70 dex (see Section  3.1).}
\label{f:fatte}
\end{figure*}

Observations and data reduction are discussed in Section~2. 
Metal abundances are presented in Section~3.
The  luminosity-metallicity relation and the radial velocities we determined 
for
the RR Lyrae stars in Sculptor are discussed in Sections~4 and 5, respectively.
In Section~6 the 
metallicity distribution 
of the Sculptor 
RR Lyrae stars 
is compared 
with that of other old stellar
populations in the galaxy. 
A summary and some final remarks in Section~7 close the paper.

\section{Observations and data  reductions}

Observations of 110 variable stars (107 RR Lyrae, 1 Anomalous Cepheid,
1 binary system, and 1 variable of unknown type) in
Sculptor and of RR Lyrae stars in 4 Galactic globular clusters 
(namely: M~2, M~15, NGC~6171, and NGC~6441) were 
carried out using FORS2 (FOcal Reducer/low dispersion 
Spectrograph 2), mounted 
on 
the ESO Very Large Telescope 
(Paranal, Chile). The data were collected in service mode during the period 
July 29 to August 5, 2003. Typical seeing values during the observations 
were in the range 0$\farcs$7--1$\farcs7$ and on average of about 1$\farcs2$.
We used 
the MXU (Mask eXchange Unit) configuration, 
that allows to observe simultaneously many objects with 
more freedom in choosing the location, size and shape of individual 
slitlets with respect to the standard MOS mode. 
The detector is a mosaic of two MIT CCDs with 15 $\mu$m pixel size.

Spectra were collected using the blue grism GRIS\_600B covering the 
3450-5900 \AA\, wavelength range, with a dispersion of 50 \AA mm$^{-1}$ with 
slits 
1$^{\prime\prime}$ wide, and usually 
14$^{\prime\prime}$ long
to allow for
sky subtraction.
With this configuration, each pixel corresponds to 0.75 \AA.
An effort was made to cover for each star the relevant wavelength range 
($\sim$ 3900-5100 \AA) containing both the CaII K and the hydrogen 
Balmer lines up to H$\beta$. 
We have used an instrumental set-up similar (i.e. same spectral range,
resolution, and typical S/N) to that employed in our study of the RR Lyrae
stars in the LMC \citep{g04}, so that RR Lyrae variables in some calibrating GCs
 are already available
(namely in clusters NGC~1851, NGC~3201, and M~68). Exposure times on the Sculptor
variables were of 31 min, as an optimal
compromise between S/N and time resolution of the light curve of the RR Lyrae
targets. 
We employed 9 masks in 
Sculptor, 2 in NGC 6441, 1 in M2, 1 in NGC 6171  and 1 in M15. 
The 9 Sculptor fields were slightly 
overlapped, so that for 25 variables we have more 
than 1 spectrum. A detailed log of the observations is given 
in Table~\ref{t:tablelog} where N is the number of variable stars observed in 
each mask.
The complete listing of the variables observed in Sculptor is provided in 
Table~\ref{t:sculptor} where we have adopted \citet{kal95} identification 
numbers.
Their location on a $16\arcmin \times 16\arcmin$ map of the
central region of Sculptor dSph galaxy is shown in Fig.~\ref{f:fatte}.
Finding charts corresponding to  the nine $6.8\arcmin \times 6.8 \arcmin$ 
FORS2 subfields
are given in  Appendix A. 
Centre of field coordinates are provided in Table~\ref{t:tablelog}. Equatorial
coordinates for all our targets can be found in table~2 of \citet{kal95}.  

\begin{table}
\begin{center}
\caption[]{Log of the observations.}
\scriptsize
\begin{tabular}{lcccrr}
\hline
\noalign{\smallskip}
Field  & RA & DEC & Date & Exptime & N\\
       &(JD2000) &(JD2000) & (UT)& (sec)\\
\noalign{\smallskip}
\hline
\noalign{\smallskip}
M15        & 21:30:13.4 &   12:11:50.4 & 2003-07-31 &   150 & 9  \\
M2         & 21:33:40.4 &  -00:50:45.7 & 2003-08-05 &   150 & 6  \\
NGC6171    & 16:32:34.6 &  -13:06:46.8 & 2003-07-29 &   150 & 6  \\
NGC6441A   & 17:50:19.9 &  -37:05:21.9 & 2003-07-29 &   300 & 5  \\
NGC6441B   & 17:50:24.5 &  -37:00:06.3 & 2003-07-30 &   300 & 7  \\
SCL1	   & 01:00:30.7 &  -33:38:18.4 & 2003-08-01 &  1860 & 15 \\
SCL2	   & 01:00:08.7 &  -33:38:18.5 & 2003-08-01 &  1860 & 18 \\
SCL3	   & 00:59:51.2 &  -33:38:18.5 & 2003-08-02 &  1860 & 15 \\
SCL4	   & 01:00:34.8 &  -33:42:43.1 & 2003-08-02 &  1860 & 15 \\
SCL5	   & 01:00:10.3 &  -33:42:43.5 & 2003-08-02 &  1860 & 19 \\
SCL6	   & 00:59:49.1 &  -33:42:45.0 & 2003-08-03 &  1860 & 18 \\
SCL7	   & 01:00:32.0 &  -33:47:18.5 & 2003-08-03 &  1860 & 15 \\
SCL8	   & 01:00:09.5 &  -33:47:15.5 & 2003-08-05 &  1860 & 19 \\
SCL9	   & 00:59:47.8 &  -33:47:53.8 & 2003-08-05 &  1860 & 13 \\
\noalign{\smallskip}
\hline
\normalsize
\end{tabular}
\label{t:tablelog}
\end{center}
\end{table}

\begin{figure} 
\includegraphics[width=8.8cm]{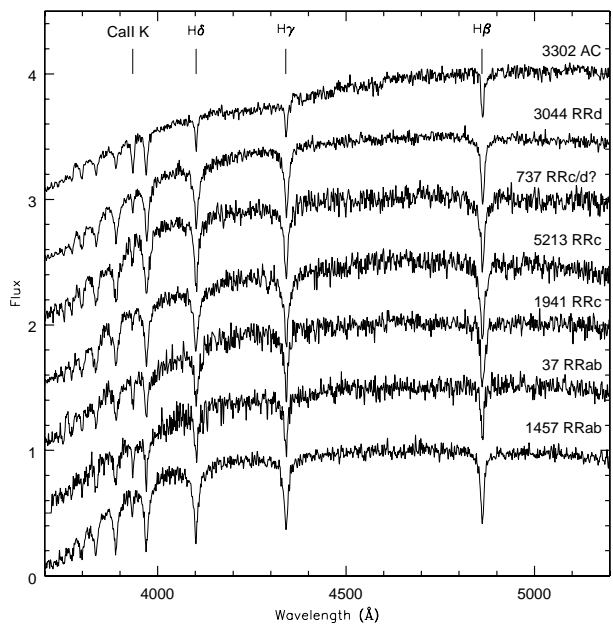}
\caption[]{Examples of spectra of variable stars in Sculptor
obtained using FORS2. 
The stars are identified according to \citet{kal95}. 
The six lower plots are RR Lyrae variables 
(from bottom to top: two fundamental mode, two first overtone, one suspected and
one confirmed double-mode pulsator observed at different phases). 
The upper plot is an Anomalous 
Cepheid (star 3302). The spectra 
have been offset for clarity, and the main spectral lines
are marked.}
\label{f:figspectra}
\end{figure}

Data reduction was performed using the standard IRAF\footnote{
IRAF is distributed by the NOAO, which are operated by AURA, under contract
with NSF} 
routines. Images 
have been trimmed, corrected for bias and for the normalized flat field. Then 
we used the IRAF command {\it lineclean} to reduce the contamination 
by cosmic rays. Up to 19 spectra were present in each pointing, and 
were extracted with the optimal extraction and automated cleaning options 
switched on. The sky contribution was subtracted making use of the slit length.
The contamination of targets from nearby stars was reduced to a minimum, 
except for a few objects.  For each science mask  a HeCdHg lamp was acquired,
and used to calibrate in wavelength the spectra, each one covering a 
different spectral range, depending on the target position. Not less than 
10 lines of the calibration lamp were visible for each aperture, and the 
resulting dispersion solutions have r.m.s. of about 0.03 \AA. Further cleaning 
of cosmic rays hits and bad sky subtractions was done using the clipping 
option in the IRAF {\it splot} task. Fig.~\ref{f:figspectra} shows examples 
of the final spectra.

\section{Derivation of the metal abundances}

We obtained spectra for 110 of the variable stars identified 
in Sculptor by \citet{kal95}, and for 25
of them we have multiple observations. 
These authors published photometry in the $V$ band 
for all our targets. Periods and
epochs of maximum light were determined from their time
series data (kindly made available by Dr. J. Kaluzny),
 using the period search package GRaphycal Analyzer
of TIme Series (GRATIS, Di Fabrizio 1999, Clementini et al. 2000). The new
ephemerides are provided in 
Table~\ref{t:sculptor}.
We found that our periods are in general slightly different from those published
by \citet{kal95}; differences are in most cases around the fourth or fifth digit.  
However, there are a number of cases where our periods and type classifications 
significantly differ from those of \citet{kal95} who published aliases of the
periodicities preferred here.
All these objects have been flagged in the last column
of Table~\ref{t:sculptor}, where we provide  
comments on individual stars.
We used our new periods to phase the spectra  
of our target stars 
in Sculptor 
since the scatter in the light curves appears
to be significantly reduced than using the published values.
Phases corresponding to the Heliocentric Julian Day (HJD) at half exposure
 are listed in 
 Column 3 of 
 Table~\ref{t:sculptor}; for 
the double-mode RR Lyrae stars they correspond to the first overtone pulsation period. 
We also computed 
intensity-averaged luminosities for all the variables in our study,
that are given in 
Column 5 of 
Table~\ref{t:sculptor}.
Based on our study of the light curves our sample contains 62 {\it ab-}type,
40 {\it c-}type, 3 confirmed and 2 suspected {\it d-}type RR Lyrae stars, 
1 Anomalous 
Cepheid, 1 suspected binary system, and 1 variable of unknown type. 

\subsection{Metallicities and metal abundance distribution of the variable
stars in Sculptor}

Precise and homogeneous metal abundances for the target stars in Sculptor 
were measured using 
the revised version of the $\Delta S$ method \citep{p59}
devised by \citet{g04}.
We do not actually measure $\Delta$S values, but rather estimate metallicities
for individual variables by comparing the strength of the H lines and of the 
K Ca II
line with analogous data for variables in GCs of known metallicity. 
A detailed description of 
our method can be found in \citet{g04}. A summary of the 
technique, and an update
of the calibration procedures are provided in 
Appendix B, to which the interested reader is referred to for details.
Spectral line indices measured for the variables in Sculptor 
following \citet{g04}, 
are given in Columns from 8 to 12 of 
Table~\ref{t:sculptor}.
\begin{figure} 
\includegraphics[width=8.8cm]{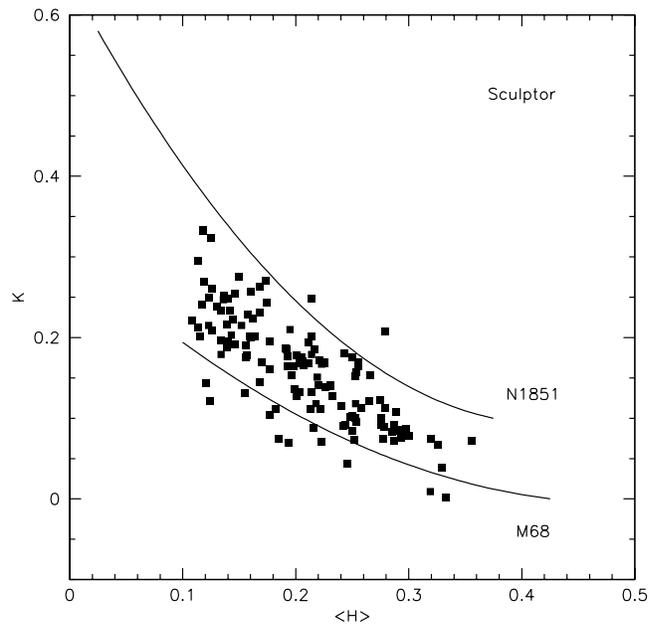}
\caption[]{Correlation between $K$ and $\langle H\rangle$ spectral indices for the
variable stars in Sculptor. Solid lines represent the mean lines for
M\,68 and NGC\,1851 from \citet{g04}.}
\label{f:fig2}
\end{figure}
The correlation between $K$ and $\langle H\rangle$ spectral indices 
is shown in Fig.~\ref{f:fig2}. 
In this figure the solid lines 
represent the mean relations of the calibrating clusters M\,68 and NGC\,1851
(see Appendix B).
The variables in Sculptor fall almost entirely below the mean line of NGC\,1851
indicating that they are metal-poorer than this cluster
([Fe/H]$_{ZW}$=$-$1.36), and also extend below the mean line of M\,68, 
showing that there is a number of variables in Sculptor metal-poorer than
[Fe/H]$_{ZW}$=$-$2.09. 
Metallicity indices $M.I.$ defined according to equations 
(1), (2) and (3) in Appendix B, are listed in Column 13 of
Table~\ref{t:sculptor}.
Individual metallicities were derived 
from the calibration equations (4) and (5) in Appendix B 
in the Zinn \& West (1984, hereafter ZW84) and in 
the Carretta \& Gratton (1997, hereafter CG97) 
metallicity scales separately; they are listed in 
Columns 14 and 15 of Table~\ref{t:sculptor}.
For different observations of the same star the [Fe/H] values 
are averages of all the available measurements. 
From the objects with multiple observations we estimate
that errors of individual abundance determinations are 
0.15 and 0.16 dex in ZW84  and CG97 metallicity scales,
respectively. 

The average metal abundance of the 107 RR Lyrae stars in our sample is 
[Fe/H]=$-$1.83$\pm$0.03 (r.m.s.=0.26 dex) with a total metallicity range
of $-2.42<$[Fe/H]$<-0.85$ in ZW84 scale, 
and an offset of about 0.2 dex to higher metallicity in CG97 scale.
The star-to-star scatter inferred from the r.m.s. dispersions is
about 0.19-0.23 dex, hence larger than typical measurement errors.
If we adopt 0.23 dex as the measured metallicity spread, and
quadratically subtract a [Fe/H] measurement error of 0.16 dex, we
obtain $\sim 0.16$ dex as our estimate of the intrinsic spread in the
metal abundance of the Sculptor RR Lyrae stars.

\begin{figure} 
\includegraphics[width=8.8cm]{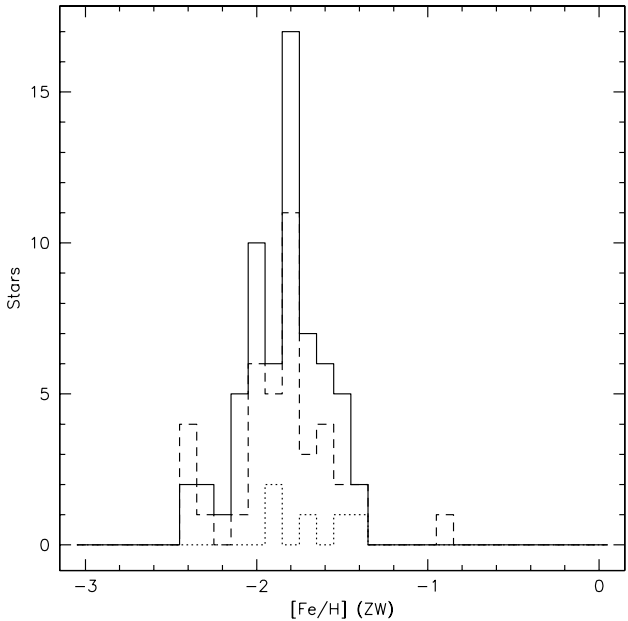}
\caption[]{Metallicity distribution of the RR Lyrae stars in 
Sculptor using metal abundances in ZW84 metallicity 
scale. Variables are divided by type, solid line: RRab pulsators,
dashed line: RRc pulsators, dotted line: RRd pulsators.}
\label{f:fig3hbw}
\end{figure}

The observed 
metallicity distribution of the Sculptor RR Lyrae stars 
(uncorrected for the measurement errors) in the 
ZW84 metallicity scale is 
shown in Fig.~\ref{f:fig3hbw}.
Variable stars are divided by type. We find that 
the different types follow the same metallicity distributions. 
As well known, CG97 scale provides systematically higher 
metallicities than ZW84. In the following we will adopt ZW84 values, unless
explicitly noted.
However, independently of the adopted metallicity scale 
there are only very few metal-rich
stars in our RR Lyrae sample.
Based on the period distribution of the {\it ab-}type pulsators,  
\cite{kal95} conclude that the bulk of the RR Lyrae stars in Sculptor
have metallicities equal to or lower than [Fe/H]$= -1.7$.
Indeed, there are only 26 stars with 
[Fe/H]$> -1.7$ in our sample.
This is also the metallicity at which 
 \citet{tol04} find the break between 
metal-poor 
and metal-rich old 
populations in Sculptor. We will come back to this point in Section 6.

   \begin{table*}
   \fontsize{7}{7}
      \caption[]{Line indices and metal abundances of  
      variable stars in Sculptor.}
         \label{t:sculptor}
     $$
         \begin{array}{rcccccccccccrccclp{0.5\linewidth}ccccccccccccccccc}
            \hline
            \noalign{\smallskip}
            {\rm Star}  &  {\rm Type}  & {\rm P} & {\rm Epoch}      & \langle V\rangle & {\rm HJD}
	            & \phi &  K 
	    &  {\rm H}_{\delta}  &  {\rm H}_{\gamma}  &  {\rm H}_{\beta}  &  \langle H\rangle  &  M.I.  &  
	    {\rm [Fe/H]}  &  {\rm [Fe/H]}  & ~V_{r}& {\rm Notes}\\
            {\rm (a)}   &  {\rm (b)}      & {\rm (b)}       & {\rm (b)} &  {\rm (b)}   & {\rm (c)}
	     &       &   
		   &             &     	       &              &    &    &  {\rm ZW} 		   &  
		   {\rm CG}  & {\rm km s^{-1}}  & \\
            \noalign{\smallskip}
            \hline
            \noalign{\smallskip}
33    &  {\rm c}	 &   0.309071  & 9168.9087  & 20.110  & 6.7675  &    0.088 &   0.002  & 0.352 & 0.347 & 0.300  & 0.333 & -0.280  & -2.42   & -2.26  & ~89  &\\
37    &  {\rm ab}	  &   0.508709  & 9162.9151  & 20.189  & 6.7675  &    0.230 &	0.161  & 0.168 & 0.199 & 0.164  & 0.177 &  0.239  & -1.96   & -1.78  & ~68  &\\
59    &  {\rm c}       &   0.359681  & 9168.9359  & 20.176  & 6.7675  &    0.073 &   0.073  & 0.277 &
0.253 & 0.226  & 0.252 &  0.035  & -2.14   & -1.97  & 111  &{\rm (d)}\\
71    &  {\rm c}       &   0.434922  & 9228.8270  & 20.015  & 6.7675  &    0.589 &   0.104  & 0.196 & 0.159 & 0.176  & 0.177 & -0.127  & -2.28   & -2.12  & 105  &\\
321   &  {\rm ab}      &   0.601783  & 9167.9003  & 20.137  & 4.7959  &   0.616 &   0.234  & 0.120 & 0.133 & 0.148  & 0.134 &  0.386  & -1.77	& -1.57  & 101  &\\
      &          &             &            &         & 6.7675  &    0.892 &   0.224  & 0.172 & 0.141 & 0.172  & 0.162 &  0.528  &         &        & ~99  &\\
368   &  {\rm c}       &   0.358962  & 9184.8696  & 20.133  & 4.7959  &   0.727 &   0.096  & 0.269 & 0.249 & 0.245  & 0.254 &  0.252  & -1.95	& -1.76  & 108  &\\
439   &  {\rm ab}      &   0.496115  & 8809.6429  & 20.146  & 4.7959  &    0.658 &   0.215  & 0.141 & 0.161 & 0.152  & 0.152 &  0.401  & -1.81   & -1.62  & 109  &\\
737   &  {\rm c/d?}    &   0.338614  & 8809.8469  & 20.061  & 4.7959  &    0.612 &   0.083  & 0.308 &
0.294 & 0.252  & 0.285 &  0.328  & -1.88   & -1.69  & ~96  &{\rm (e)}\\
763   &  {\rm c}       &   0.293114  & 8830.8675  & 20.139  & 3.7485  &    0.615 &   0.128  & 0.280 & 0.239 & 0.176  & 0.232 &  0.370  & -1.84   & -1.65  & 102  &\\
786   &  {\rm c}       &   0.279088  & 8809.7457  & 20.179  & 3.7485  &    0.048 &   0.072  & 0.362 & 0.377 & 0.328  & 0.356 &  0.601  & -1.64   & -1.43  & 111  &\\
793   &  {\rm c}       &   0.359921  & 8809.9129  & 20.135  & 3.7485  &    0.339 &   0.118  & 0.268 & 0.264 & 0.228  & 0.253 &  0.442  & -1.91   & -1.72  & 107  &\\
      &  	 &             &            &         & 4.7959  &    0.249 &   0.092  & 0.253 & 0.250 & 0.230  & 0.244 &  0.150  &	   &	    & 140  &\\
803   &  {\rm ab}      &   0.573658  & 8809.8060  & 20.061  & 3.7485  &    0.394 &   0.222  & 0.135 & 0.131 & 0.167  & 0.144 &  0.392  & -1.82   & -1.63  & ~92  &\\
811   &  {\rm c}       &   0.363134  & 8809.8517  & 20.131  & 3.7485  &    0.098 &   0.092  & 0.308 & 0.283 & 0.268  & 0.287 &  0.420  & -1.93   & -1.75  & 131  &\\
      &  	 &             &            &         & 4.7959  &    0.982 &   0.090  & 0.265 & 0.229 & 0.232  & 0.242 &  0.120  &	   &	    & 140  &\\
860   &  {\rm c}       &   0.359517  & 9158.9314  & 20.206  & 4.7959  &    0.071 &   0.140  & 0.299 &
0.208 & 0.185  & 0.231 &  0.464  & -1.76   & -1.56  & 129  &{\rm (f)}\\
1139  &  {\rm ab}      &   0.518930  & 8809.5982  & 20.060  & 3.7485  &    0.245 &   0.075  & 0.299 & 0.262 & 0.269  & 0.277 &  0.194  & -2.00   & -1.82  & 127  &\\
1164  &  {\rm ab}      &   0.606190  & 8809.6323  & 20.185  & 3.7485  &    0.367 &   0.261  & 0.152 & 0.113 & 0.113  & 0.126 &  0.473  & -1.75   & -1.56  & 126  &\\
1261  &  {\rm ab}      &   0.630359  & 8809.4200  & 20.172  & 3.7485  &    0.912 &   0.323  & 0.106 & 0.134 & 0.137  & 0.125 &  0.782  & -1.48   & -1.26  & 148  &\\
1439  &  {\rm c}       &   0.356043  & 9168.9196  & 20.152  & 6.8001  &    0.952 &   0.115  & 0.292 &
0.230 & 0.199  & 0.240 &  0.324  & -1.88   & -1.70  & 104  &{\rm (d)}\\
1457  &  {\rm ab}      &   0.717799  & 8827.8527  & 19.919  & 6.8001  &    0.915 &   0.085  & 0.341 & 0.295 & 0.243  & 0.293 &  0.398  & -1.82   & -1.63  & ~74  &\\
1491  &  {\rm c}       &   0.357866  & 8809.7866  & 20.135  & 6.8001  &    0.727 &   0.153  & 0.350 &
0.225 & 0.223  & 0.266 &  0.854  & -1.41   & -1.20  & ~84  &{\rm (g)}\\
1519  &  {\rm c}       &   0.356859  & 9169.9083  & 20.173  & 6.7675  &    0.414 &   0.127  & 0.228 & 0.193 & 0.182  & 0.201 &  0.161  & -2.03   & -1.85  & ~79  &\\
1546  &  {\rm ab}      &   0.531239  & 9174.9206  & 20.151  & 6.8001  &    0.737 &   0.143  & 0.076 & 0.158 & 0.129  & 0.121 & -0.153  & -2.31   & -2.14  & ~92  &\\
1555  &  {\rm ab}      &   0.527243  & 8823.8298  & 20.089  & 6.7675  &    0.112 &   0.078  & 0.312 & 0.301 & 0.286  & 0.300 &  0.372  & -1.84   & -1.65  & ~81  &\\
1558  &  {\rm c/d?/Bl} &   0.243016  & 8833.9116  & 20.203  & 6.8001  &    0.006 &   0.102  & 0.296 &
0.250 & 0.198  & 0.248 &  0.267  & -1.65   & -1.45  & ~94  &{\rm (h)}\\
      &          &             &            &         & 6.7675  &    0.872 &   0.175  & 0.259 & 0.265 & 0.225  & 0.250 &  0.913  &	   &	    & 107  &\\
1566  &  {\rm ab}      &   0.570272  & 8809.9982  & 20.053  & 6.8001  &    0.266 &   0.177  & 0.175 & 0.210 & 0.193  & 0.193 &  0.458  & -1.77   & -1.57  & 107  &\\
1823  &  {\rm c}      &   0.298462  & 9189.8649  & 20.255  & 4.7959  &    0.373 &   0.152  & 0.294 & 0.230 & 0.235  & 0.253 &  0.743  & -1.52   & -1.30  & 107  &\\
1830  &  {\rm ab}      &   0.517855  & 8809.2712  & 20.224  & 3.8132  &    0.181 &   0.145  & 0.180 & 0.165 & 0.160  & 0.168 &  0.089  & -2.09   & -1.92  & 112  &\\
1875  &  {\rm ab}      &   0.499773  & 8823.9437  & 20.028  & 3.8132  &    0.400 &   0.188  & 0.153 &
0.151 & 0.113  & 0.139 &  0.174  & -2.02   & -1.84  & ~95  &{\rm (f)}\\
1899  &  {\rm ab}      &   0.646664  & 9182.9074  & 19.978  & 4.7959  &    0.201 &   0.071  & 0.240 & 0.219 & 0.210  & 0.223 & -0.137  & -2.29   & -2.13  & ~74  &\\
1910  &  {\rm ab}      &   0.572828  & 9189.9080  & 20.083  & 3.8132  &    0.172 &   0.111  & 0.183 & 0.185 & 0.177  & 0.182 & -0.059  & -2.11   & -1.93  & ~97  &\\
      &          &             &            &         & 6.8001  &    0.386 &   0.209  & 0.133 & 0.128 & 0.118  & 0.126 &  0.208  &	   &	    & ~77  &\\
1914  &  {\rm ab}      &   0.570540  & 8829.8781  & 20.220  & 6.8001  &    0.084 &   0.195  & 0.120 & 0.106 & 0.194  & 0.140 &  0.213  & -1.87   & -1.68  & ~98  &\\
      &   	 &             &            &         & 6.7675  &    0.027 &   0.248  & 0.137 & 0.133 & 0.138  & 0.136 &  0.478  &	   &	    & ~78  &\\
1930  &  {\rm ab}      &   0.611160  & 9188.9162  & 20.251  & 3.8132  &    0.626 &   0.141  & 0.251 & 0.235 & 0.206  & 0.231 &  0.473  & -1.79   & -1.59  & 103  &\\
      &  	 &             &            &         & 6.8001  &    0.514 &   0.165  & 0.220 & 0.170 &
      0.190  & 0.193 &  0.380  &	   &	    & 119  &{\rm (g)}\\
1932  &  {\rm ab}      &   0.506044  & 9174.9206  & 20.155  & 4.7959  &    0.848 &   0.275  & 0.143 & 0.152 & 0.155  & 0.150 &  0.731  & -1.53   & -1.32  & 118  &\\
      &   	 &             &            &         & 6.8001  &    0.809 &   0.257  & 0.228 & 0.141 & 0.113  & 0.160 &  0.716  &	   &	    & ~90  &\\
1940  &  {\rm ab}      &   0.692975  & 9224.7748  & 20.117  & 4.7959  &    0.312 &   0.332  & 0.127 & 0.137 & 0.091  & 0.118 &  0.769  & -1.49   & -1.28  & ~98  &\\
      &  	 &             &            &         & 6.7675  &    0.157 &   0.333  & 0.109 & 0.126 & 0.117  & 0.118 &  0.769  &	   &	    & 104  &\\
1941  &  {\rm c}       &   0.365674  & 9166.8670  & 20.152  & 6.7675  &    0.689 &   0.157  & 0.258 & 0.253 & 0.250  & 0.254 &  0.789  & -1.47   & -1.26  & 107  &\\
1943  &  {\rm ab}      &   0.551149  & 9169.9100  & 20.146  & 3.8132  &    0.045 &   0.203  & 0.172 & 0.131 & 0.127  & 0.143 &  0.281  & -1.54   & -1.32  & ~89  &\\
      &          &             &            &         & 6.8001  &    0.464 &   0.248  & 0.235 & 0.222 &
      0.184  & 0.214 &  1.159  &	   &	    & 103  &{\rm (i)}\\
1997  &  {\rm ab}      &   0.626766  & 8823.9437  & 20.136  & 3.8132  &    0.625 &   0.179  & 0.165 & 0.101 & 0.138  & 0.134 &  0.097  & -2.08   & -1.91  & 125  &\\
2058  &  {\rm ab}      &   0.503415  & 9226.9053  & 20.238  & 3.8132  &    0.610 &   0.133  & 0.224 & 0.212 & 0.204  & 0.214 &  0.288  & -1.92   & -1.73  & ~96  &\\
2423  &  {\rm c}       &   0.358540  & 9167.9003  & 20.010  & 3.8132  &    0.328 &   0.202  & 0.170 &
0.149 & 0.157  & 0.159 &  0.372  & -1.84   & -1.65  & 120  &{\rm (f)}\\
2424  &  {\rm c}       &   0.348780  & 8809.8058  & 20.145  & 4.7959  &    0.527 &   0.166  & 0.215 & 0.227 & 0.178  & 0.207 &  0.482  & -1.75   & -1.55  & 118  &\\
            \noalign{\smallskip}
            \hline
         \end{array}
     $$
   \end{table*}


   \begin{table*}
   \fontsize{7}{7}
   \addtocounter{table}{-1}
      \caption[]{Cont.}
     $$
         \begin{array}{rcccccccccccrccclp{0.5\linewidth}ccccccccccccccccc}
            \hline
            \noalign{\smallskip}
            {\rm Star}  &  {\rm Type}  & {\rm P} & {\rm Epoch}      & \langle V\rangle & {\rm HJD}
	            & \phi &  K 
	    &  {\rm H}_{\delta}  &  {\rm H}_{\gamma}  &  {\rm H}_{\beta}  &  \langle H\rangle  &  M.I.  &  
	    {\rm [Fe/H]}  &  {\rm [Fe/H]}  & ~V_{r}& {\rm Notes}\\
            {\rm (a)}   &  {\rm (b)}      & {\rm (b)}       & {\rm (b)} &  {\rm (b)}   & {\rm (c)}
	     &       &   
		   &             &     	       &              &    &    &  {\rm ZW} 		   &  
		   {\rm CG}  & {\rm km s^{-1}}  & \\
            \noalign{\smallskip}
            \hline
            \noalign{\smallskip}
2450	& {\rm ab}	  & 0.617962  &  8810.1279  &  20.071  &  2.7656 &   0.886  &	0.252  &  0.141  &  0.141  &   0.126  &  0.136   &   0.500   &  -1.84  &   -1.65 & 113 &\\
        &	  &           &             &          &  3.8132 &   0.581  &   0.238  &  0.123  &  0.135  &   0.133  &  0.130	 &   0.383   &	       &         & ~92 &\\
        &	  &           &             &          &  4.7959 &   0.172  &   0.192  &  0.136  &  0.157  &   0.146  &  0.146	 &   0.239   &	       &	 & 108 &\\
2455    & {\rm ab}	& 0.636250  &  9184.8700  &  20.156  &  3.7485 &   0.413  &   0.247  &  0.160  &  0.135  &   0.112  &  0.136   &   0.467   &  -1.76  &   -1.56 & 113 &\\
2458    & {\rm c}	& 0.357686  &  8809.6823  &  20.156  &  2.7656 &   0.446  &   0.118  &  0.260  &  0.208  &   0.187  &  0.218   &   0.199   &  -1.73  &   -1.53 & ~85 &\\
        &         &           &             &          &  3.8132 &   0.375  &	0.201  &  0.228  &  0.208  &   0.205  &  0.214   &   0.802   &         &	 & 102 &\\
2470	& {\rm ab}	& 0.693447  &  8809.4598  &  20.073  &  3.8132 &   0.245  &   0.168  &  0.183  &  0.220  &   0.207  &  0.203   &   0.467   &  -1.76  &   -1.56 & ~94 &\\
2482    & {\rm c}	& 0.365865  &  9167.8931  &  20.191  &  4.7959 &   0.233  &   0.108  &  0.337  &
 0.250  &   0.279  &  0.289   &   0.600   &  -1.64  &   -1.44 & 135 &{\rm (f)}\\
2502    & {\rm ab}	& 0.487474  &  9185.8606  &  20.238  &  4.7959 &   0.419  &   0.123  &  0.268  &
 0.300  &   0.258  &  0.275   &   0.645   &  -1.60  &   -1.39 & 140 &{\rm (f)}\\
2545    & {\rm ab}	& 0.674059  &  8809.0126  &  20.016  &  2.7656 &   0.104  &   0.168  &  0.224  &  0.231  &   0.220  &  0.225   &   0.644   &  -1.60  &   -1.39 & 116 &\\
2566    & {\rm ab}	& 0.583517  &  9190.8800  &  20.234  &  4.7959 &   0.018  &   0.180  &  0.248  &  0.235  &   0.247  &  0.243   &   0.901   &  -1.38  &   -1.15 & ~88 &\\
2627    & {\rm ab}	& 0.575091  &  9166.8885  &  20.130  &  4.7959 &   0.733  &   0.192  &  0.136  &  0.157  &   0.146  &  0.146   &   0.239   &  -1.96  &   -1.78 & 108 &\\
2689    & {\rm ab}	& 0.511352  &  8809.5668  &  20.139  &  3.7485 &   0.805  &   0.270  &  0.155  &  0.206  &   0.159  &  0.173   &   0.909   &  -1.37  &   -1.15 & ~99 &\\
3004    & {\rm ab}	& 0.715496  &  8809.5055  &  20.044  &  2.7656 &   0.990  &   0.187  &  0.199  &  0.183  &   0.192  &  0.191   &   0.514   &  -1.72  &   -1.52 & 129 &\\
3016    & {\rm c}	& 0.360325  &  8809.8443  &  20.117  &  2.7656 &   0.220  &   0.139  &  0.279  &
 0.227  &   0.172  &  0.226   &   0.414   &  -1.81  &   -1.61 & 131 &{\rm (d)}\\
3019    & {\rm ab}	& 0.732965  &  9162.9250  &  19.978  &  3.7485 &   0.472  &   0.202  &  0.112  &  0.094  &   0.138  &  0.115   &   0.109   &  -2.07  &   -1.90 & 105 &\\
3024    & {\rm ab/Bl?}  & 0.572818  &  9223.7730  &  20.059  &  3.7485 &   0.052  &   0.179  &  0.197  &
 0.229  &   0.216  &  0.214   &   0.641   &  -1.61  &   -1.40 & 143 &{\rm (f)}\\
3043    & {\rm ab}	& 0.623908  &  8810.0604  &  20.192  &  2.7656 &   0.650  &   0.178  &  0.201  &  0.199  &   0.204  &  0.201   &   0.529   &  -1.70  &   -1.50 & 129 &\\
3044    & {\rm d}	& 0.354271  &  8810.0791  &  20.126  &  3.7485 &   0.044  &   0.092  &  0.319  &
 0.283  &   0.227  &  0.276   &   0.355   &  -1.86  &   -1.67 & 133 &{\rm (l)}\\
3104    & {\rm d}	& 0.356984  &  8809.9225  &  20.209  &  2.7656 &   0.982  &   0.243  &  0.182  &
 0.171  &   0.169  &  0.174   &   0.745   &  -1.51  &   -1.30 & 124 &{\rm (l)}\\
3113	& {\rm ab}	& 0.593260  &  8809.6681  &  20.098  &  2.7656 &   0.049  &   0.121  &  0.296  &  0.244  &   0.256  &  0.265   &   0.556   &  -1.79  &   -1.59 & 130 &\\
	& 	  &           &             &          &  3.7485 &   0.706  &	0.216  &  0.152  &  0.134  &   0.130  &  0.139   &   0.319   &         &	 & 116 &\\
3125    & {\rm ab}	& 0.533130  &  8809.9264  &  20.120  &  2.7656 &   0.220  &   0.201  &  0.179  &  0.160  &   0.154  &  0.164   &   0.404   &  -1.81  &   -1.62 & 134 &\\
3143    & {\rm d}	& 0.354531  &  9184.9270  &  20.188  &  2.7656 &   0.607  &   0.170  &  0.276  &
 0.236  &   0.251  &  0.255   &   0.914   &  -1.36  &   -1.14 & 140 &{\rm (l)}\\
3302	& {\rm AC}	& 1.346056  &  9227.8700  &  18.560  &  4.8256 &   0.506  &   0.295  &  0.101  &  0.116  &   0.124  &  0.114   &   0.553   &  -1.78  &   -1.58 & ~81 &\\
	& 	  &           &             &          &  6.8001 &   0.975  &	0.089  &  0.289  &  0.294  &   0.251  &  0.278   &   0.340   &         &	 & ~84 &\\
3318    & {\rm ab}	& 0.640243  &  9169.8898  &  20.148  &  4.8256 &   0.530  &   0.221  &  0.086  &  0.108  &   0.131  &  0.108   &   0.165   &  -2.02  &   -1.85 & ~79 &\\
3319    & {\rm ab/Bl}	& 0.564984  &  9223.8400  &  20.012  &  4.8256 &   0.703  &   0.229  &  0.172  &  0.148  &   0.153  &  0.158   &   0.524   &  -1.71  &   -1.51 & ~89 &\\
3320	& {\rm c}	& 0.282469  &  9192.9040  &  20.119  &  4.8256 &   0.954  &   0.067  &  0.341  &  0.338  &   0.300  &  0.326   &   0.403   &  -1.94  &   -1.76 & ~84 &\\
	& 	  &           &             &          &  6.8001 &   0.944  &	0.039  &  0.358  &  0.338  &   0.290  &  0.329   &   0.115   &         &	 & ~66 &\\
3346	& {\rm c}	& 0.357547  &  9168.9247  &  20.125  &  4.8256 &   0.857  &   0.088  &  0.244  &  0.216  &   0.189  &  0.216   &  -0.043   &  -2.02  &   -1.84 & ~96 &\\
	& 	  &           &             &          &  6.8001 &   0.379  &	0.141  &  0.214  &  0.243  &   0.203  &  0.220   &   0.394   &         &	 & 120 &\\
3365    & {\rm ab}	& 0.668088  &  9235.8929  &  20.105  &  6.8001 &   0.807  &   0.196  &  0.126  &  0.157  &   0.119  &  0.134   &   0.187   &  -2.01  &   -1.82 & ~99 &\\
3413	& {\rm c}	& 0.359537  &  9167.9120  &  20.132  &  4.8256 &   0.607  &   0.194  &  0.247  &  0.206  &   0.181  &  0.211   &   0.727   &  -1.72  &   -1.52 & ~90 &\\
	& 	  &           &             &          &  6.8001 &   0.098  &	0.103  &  0.299  &  0.230  &   0.220  &  0.250   &   0.285   &         &	 & ~85 &\\
3710	& {\rm Bin?}	& 0.473839  &  9192.9030  &  19.856  &  3.8132 &   0.070  &   0.046  &  0.215  &
 0.205  &   0.182  &  0.201   &  -0.420   &  -2.56  &   -2.42 & ~73 &{\rm (f)}\\
        &         &           &             &          &  4.8256 &   0.207  &   0.077  &  0.201  &  0.159  &   0.194  &  0.185   &  -0.270   &	       &	 & ~85 &\\
        &         &           &             &          &  6.8001 &   0.374  &   0.029  &  0.159  &  0.186  &   0.140  &  0.162   &  -0.646   &	       &	 & ~89 &\\
3827    & {\rm ab}	& 0.587708  &  9226.7339  &  20.135  &  6.8001 &   0.644  &   0.212  &  0.147  &  0.071  &   0.122  &  0.114   &   0.152   &  -2.04  &   -1.86 & ~67 &\\
3834	& {\rm c}	& 0.375237  &  9189.9150  &  20.003  &  3.7809 &   0.125  &   0.169  &  0.179  &  0.158  &   0.172  &  0.170   &   0.245   &  -1.84  &   -1.65 & ~96 &\\
        &  	  &           &             &          &  4.8256 &   0.909  &	0.164  &  0.204  &  0.205  &   0.184  &  0.198   &   0.400   &         &	 & ~85 &\\
        &  	  &           &             &          &  6.8001 &   0.171  &	0.195  &  0.184  &  0.164  &   0.184  &  0.177   &   0.463   &         &	 & ~90 &\\
3862	& {\rm c}	& 0.294092  &  9188.9188  &  20.219  &  3.7809 &   0.622  &   0.087  &  0.307  &  0.318  &   0.268  &  0.298   &   0.442   &  -1.78  &   -1.58 & 121 &\\
3907    & {\rm ab}	& 0.583204  &  9169.9000  &  20.203  &  3.8132 &   0.683  &   0.269  &  0.126  &  0.093  &   0.137  &  0.119   &   0.464   &  -1.76  &   -1.56 & 112 &\\
4263    & {\rm c}	& 0.284768  &  8820.7399  &  20.229  &  2.7333 &   0.861  &   0.208  &  0.284  &  0.283  &   0.268  &  0.279   &   1.493   &  -0.85  &   -0.60 & 111 &\\
4277	& {\rm c}	& 0.306303  &  8809.7972  &  20.172  &  2.7333 &   0.136  &   0.113  &  0.332  &  0.254  &   0.251  &  0.279   &   0.576   &  -1.79  &   -1.60 & 115 &\\
	& 	  &           &             &          &  3.8132 &   0.661  &	0.100  &  0.245  &  0.277  &   0.239  &  0.253   &   0.284   &         &	 & ~96 &\\
4291	& {\rm c}	& 0.387992  &  8810.1654  &  20.057  &  2.7656 &   0.282  &   0.185  &  0.209  &  0.238  &   0.202  &  0.217   &   0.700   &  -1.64  &   -1.43 & 105 &\\
	& 	  &           &             &          &  3.8132 &   0.982  &	0.186  &  0.206  &  0.191  &   0.180  &  0.192   &   0.514   &         &	 & 106 &\\
4308	& {\rm c}	& 0.358956  &  8809.7147  &  20.123  &  2.7656 &   0.364  &   0.112  &  0.229  &  0.240  &   0.170  &  0.213   &   0.122   &  -1.83  &   -1.64 & 118 &\\
	& 	  &           &             &          &  3.8132 &   0.283  &	0.169  &  0.230  &  0.220  &   0.226  &  0.225   &   0.649   &	       &	 & 130 &\\
            \noalign{\smallskip}
            \hline
         \end{array}
     $$
   \end{table*}

   \begin{table*}
   \fontsize{7}{7}
   \addtocounter{table}{-1}
      \caption[]{Cont.}
     $$
         \begin{array}{rcccccccccccrccclp{0.5\linewidth}ccccccccccccccccc}
            \hline
            \noalign{\smallskip}
            {\rm Star}  &  {\rm Type}  & {\rm P} & {\rm Epoch}      & \langle V\rangle & {\rm HJD}
	            & \phi &  K 
	    &  {\rm H}_{\delta}  &  {\rm H}_{\gamma}  &  {\rm H}_{\beta}  &  \langle H\rangle  &  M.I.  &  
	    {\rm [Fe/H]}  &  {\rm [Fe/H]}  & ~V_{r}& {\rm Notes}\\
            {\rm (a)}   &  {\rm (b)}      & {\rm (b)}       & {\rm (b)} &  {\rm (b)}   & {\rm (c)}
	     &       &   
		   &             &     	       &              &    &    &  {\rm ZW} 		   &  
		   {\rm CG}  & {\rm km s^{-1}}  & \\
            \noalign{\smallskip}
            \hline
            \noalign{\smallskip}
4313 &  {\rm ab}	 &  0.731064  & 8820.6952  &   20.111   & 2.7333   &  0.298   &   0.231 &  0.202  &  0.150  &  0.152   &   0.168  &  0.617 & -1.63  & -1.42  &  119&\\
4385 &  {\rm ab}       &  0.487413  & 8809.8949  &   20.191   & 2.7656   &  0.550   &	0.200 &  0.108  &  0.190  &  0.183   &   0.160  &  0.376 & -1.87  & -1.68  &  117&\\
     &  	 &            &            &    	& 3.8132   &  0.699   &   0.076 &  0.328  &  0.298  &  0.253   &   0.293  &  0.311 &	    &	     &  103&\\
4689 &  {\rm ab}       &  0.639202  & 8809.9119  &   20.024   & 2.7656   &  0.843   &	0.178 &  0.150  &  0.137  &  0.183   &   0.157  &  0.221 & -1.98  & -1.79  &  108&\\
4747 &  {\rm ab}       &  0.591979  & 9166.9255  &   20.181   & 2.7333   &  0.246   &	0.168 &  0.234  &  0.201  &  0.198   &   0.211  &  0.534 & -1.70  & -1.50  &  120&\\
4780 &  {\rm ?}        &  0.391270  & 8810.1007  &   19.958   & 2.7656   &  0.164   &	0.252 &  0.061 
&  0.081  &  0.103   &   0.082  &  0.160 & -2.03  & -1.85  &  110&{\rm (m)}\\
4785 &  {\rm ab}       &  0.506109  & 8809.4911  &   20.139   & 2.7656   &  0.935   &	0.069 &  0.267  &  0.179  &  0.136   &   0.194  & -0.283 & -2.42  & -2.27  &  ~63&\\
4786 &  {\rm ab}       &  0.537290  & 9226.9054  &   20.102   & 2.7333   &  0.365   &	0.263 &  0.146  &  0.172  &  0.185   &   0.168  &  0.817 & -1.45  & -1.23  &  138&\\
4793 &  {\rm ab}       &  0.559834  & 8809.1465  &   20.059   & 2.7333   &  0.839   &	0.131 &  0.163  &  0.158  &  0.142   &   0.155  & -0.068 & -2.23  & -2.06  &  133&\\
5000 &  {\rm c}        &  0.323261  & 9166.8850  &   20.158   & 4.8256   &  0.543   &	0.136 &  0.192  &  0.222  &  0.182   &   0.199  &  0.207 & -1.99  & -1.81  &  ~93&\\
5049 &  {\rm ab}       &  0.648752  & 8825.8195  &   20.076   & 4.8256   &  0.392   &	0.234 &  0.151  &  0.115  &  0.160   &   0.142  &  0.444 & -1.78  & -1.58  &  ~96&\\
5065 &  {\rm c}        &  0.380212  & 9222.7846  &   19.989   & 3.7809   &  0.931   &	0.101 &  0.322  &  0.275  &  0.231   &   0.276  &  0.440 & -1.74  & -1.54  &  125&\\
     &  	 &            &            &     	& 4.8256   &  0.678   &   0.176 &  0.251  &  0.197  &  0.167   &   0.205  &  0.547 &	    &	     &  116&\\
5105 &  {\rm ab}       &  0.556682  & 9166.8930  &   20.154   & 4.8256   &  0.847   &	0.255 &  0.162  &  0.108  &  0.168   &   0.146  &  0.588 & -1.65  & -1.45  &  132&\\
5123 &  {\rm c}        &  0.325323  & 8810.9530  &   20.162   & 4.8256   &  0.327   &	0.074 &  0.325  &  0.323  &  0.312   &   0.320  &  0.441 & -1.78  & -1.59  &  ~95&\\
5155 &  {\rm ab}       &  0.567457  & 8823.8298  &   20.090   & 4.8256   &  0.611   &	0.241 &  0.081  &  0.104  &  0.167   &   0.117  &  0.316 & -1.89  & -1.70  &  108&\\
5330 &  {\rm ab}       &  0.645766  & 8809.3652  &   20.136   & 3.7809   &  0.976   &	0.210 &  0.170 
&  0.202  &  0.214   &   0.195  &  0.704 & -1.55  & -1.34  &  110&{\rm (f)}\\
5343 &  {\rm ab/Bl?}   &  0.546964  & 9164.9231  &   20.181   & 4.8256   &  0.149   &	0.190 &  0.158  &  0.147  &  0.162   &   0.156  &  0.282 & -1.92  & -1.74  &  108&\\
5359 &  {\rm ab}       &  0.670952  & 8809.7021  &   19.972   & 3.7809   &  0.375   &	0.121 &  0.108  &  0.115  &  0.150   &   0.124  & -0.247 & -2.39  & -2.23  &  ~93&\\
5364 &  {\rm c}        &  0.394203  & 9168.9185  &   19.943   & 3.7809   &  0.624   &	0.044 &  0.238 
&  0.268  &  0.231   &   0.246  & -0.248 & -2.39  & -2.23  &  135&{\rm (f)}\\
5376 &  {\rm c}        &  0.389520  & 9187.8882  &   20.079   & 3.7809   &  0.315   &	0.074 &  0.224  &  0.167  &  0.164   &   0.185  & -0.287 & -2.42  & -2.27  &  137&\\
5382 &  {\rm ab}       &  0.595937  & 9191.9233  &   20.171   & 3.7809   &  0.708   &	0.249 &  0.121  &  0.136  &  0.111   &   0.123  &  0.389 & -1.83  & -1.63  &  ~99&\\
5393 &  {\rm c}        &  0.322741  & 9166.9255  &   20.081   & 3.7809   &  0.587   &	0.153 &  0.221  &  0.185  &  0.183   &   0.196  &  0.313 & -1.89  & -1.71  &  144&\\
5492 &  {\rm ab}       &  0.528786  & 8833.9145  &   20.161   & 3.7809   &  0.065   &	0.171 &  0.222  &  0.205  &  0.190   &   0.206  &  0.515 & -1.72  & -1.52  &  115&\\
5710 &  {\rm c}        &  0.355803  & 9227.7927  &   20.132   & 3.7809   &  0.997   &	0.133 &  0.234  &  0.205  &  0.171   &   0.203  &  0.219 & -1.98  & -1.79  &  141&\\
5714 &  {\rm c}        &  0.292913  & 8809.7224  &   20.221   & 2.7333   &  0.747   &	0.113 &  0.267  &  0.280  &  0.228   &   0.258  &  0.434 & -1.79  & -1.59  &  ~84&\\
5723 &  {\rm ab}       &  0.566020  & 9166.9255  &   20.250   & 2.7333   &  0.801   &	0.248 &  0.138  &  0.135  &  0.147   &   0.140  &  0.505 & -1.73  & -1.53  &  112&\\
5724 &  {\rm ab}       &  0.498508  & 8809.8649  &   $-$      & 2.7333   &  0.400   &	0.009 &  0.326 
&  0.333  &  0.298   &   0.319  & -0.265 & -2.40  & -2.25  &  104&{\rm (f,n)}\\
5747 &  {\rm ab}       &  0.559848  & 8833.9264  &   20.040   & 2.7333   &  0.388   &	0.176 &  0.172  &  0.145  &  0.153   &   0.156  &  0.206 & -1.99  & -1.81  &  126&\\
5751 &  {\rm c}        &  0.397328  & 8809.6162  &   20.083   & 2.7333   &  0.756   &	0.112 &  0.267  &  0.191  &  0.208   &   0.222  &  0.177 & -1.99  & -1.81  &  ~97&\\
     &  	 &            &            &    	& 3.7809   &  0.393   &   0.072 &  0.340  &  0.276  &  0.245   &   0.287  &  0.233 &	    &	     &  118&\\
5773 &  {\rm ab}       &  0.508760  & 8823.9310  &   20.223   & 3.7809   &  0.931   &	0.084 &  0.272  &  0.256  &  0.223   &   0.250  &  0.116 & -2.07  & -1.89  &  ~91&\\
5802 &  {\rm ab}       &  0.514632  & 8810.9385  &   20.244   & 3.7809   &  0.788   &	0.215 &  0.107  &  0.121  &  0.142   &   0.123  &  0.221 & -1.98  & -1.79  &  150&\\
6031 &  {\rm c}        &  0.326878  & 9168.9123  &   20.124   & 2.7333   &  0.709   &	0.172 &  0.244 
&  0.216  &  0.203   &   0.221  &  0.637 & -1.61  & -1.40  &  140&{\rm (f)}\\
6048 &  {\rm ab}       &  0.626791  & 9224.8920  &   20.168   & 2.7333   &  0.964   &	0.167 &  0.229  &  0.221  &  0.220   &   0.223  &  0.623 & -1.62  & -1.41  &  117&\\
6050 &  {\rm c}        &  0.305175  & 8809.7002  &   20.121   & 2.7333   &  0.227   &	0.151 &  0.248  &  0.228  &  0.180   &   0.219  &  0.462 & -1.76  & -1.57  &  139&\\
6085 &  {\rm c}        &  0.361528  & 9235.7159  &   20.146   & 2.7333   &  0.808   &	0.165 &  0.293  &  0.243  &  0.228   &   0.255  &  0.869 & -1.40  & -1.18  &  139&\\
            \noalign{\smallskip}
            \hline
         \end{array}
     $$
\begin{list}{}{}
\item[$^{\mathrm{a}}$] Identifiers are from \citet{kal95}
\item[$^{\mathrm{b}}$] Epochs are $-$2440000. Along with types, periods, and mean magnitudes
(in intensity average) they were redetermined from the study of the light curves based on 
data from \citet{kal95}. In a  number of cases (marked with notes) they
differ significantly from values published in \citet{kal95}.
\item[$^{\mathrm{c}}$] HJDs are $-$2452850, they correspond to the HJD at half exposure.
\item[$^{\mathrm{d}}$] \citet{kov01} classifies star 59, 1439, and 
3016 as suspected double-mode RR Lyrae stars,  with 
periodicities of 0.35968/0.4837; 0.35604/0.47809; and 0.36033/0.48401 respectively. Our study of the light curves does 
not confirm these findings,
we think these stars are monoperiodic {\it c-}type variables with noisy light curves.
\item[$^{\mathrm{e}}$] We think this is a double-mode RR Lyrae, we list the first overtone period. 
\item[$^{\mathrm{f}}$] Classification and periods differ significantly from \citet{kal95} who published aliases
of the periodicities preferred here.
\item[$^{\mathrm{g}}$] Noisy spectrum.
\item[$^{\mathrm{h}}$] We think this RR Lyrae could either be a double-mode or be affected by the Blazhko effect
(Bl, Blazhko 1907).
\item[$^{\mathrm{i}}$] G band visible.
\item[$^{\mathrm{l}}$] \citet{kov01} classifies these variables as double-mode RR Lyrae stars respectively with 
periodicities of 0.35427/0.47546 ( 3044), 0.35699/0.47975 (3104), and 0.35453/0.47630 ( 3143).
\item[$^{\mathrm{m}}$] Variable star of unknown type.
\item[$^{\mathrm{n}}$] Incomplete light curve.
\end{list}
   \end{table*}


\subsection{Comparison with metallicities from the Fourier parameters of the
light curve and the pulsation equations}

Metal abundances for the {ab-}type RR Lyrae stars in Sculptor have been estimated by 
\citet{kov01} using the parameters of the Fourier decomposition of the light curves
and the \citet{jurcsik96} method. Derived metallicities are in the range from
$\sim -0.8$ to $-2.0$ dex, with an
average value of [Fe/H]$\sim -$1.5 dex in \citet{jurcsik95} metallicity scale. 
Based on the frequency analysis of \citet{kal95} data, \citet{kov01} also 
identified 15 confirmed and 3 suspected double mode pulsators in Sculptor
and estimated their metal abundance using the pulsation equations.
 He found that, similarly to the RRab's, the bulk of the double-mode RR Lyrae stars
 in Sculptor have
[Fe/H]$\sim -1.5$, with only two RRd's 
(stars 1168 and 5354 in \citealt{kal95})
at [Fe/H]=$-1.9$.
The metallicity distributions of RRab and RRd stars separately are shown
in fig. 5 of 
\citet{kov01}, and can be compared with our distribution in 
Fig.~\ref{f:fig3hbw}.
We recall that the \citet{jurcsik95} metallicity scale is on average 0.2 dex
more metal rich than the ZW84 scale at [Fe/H]$\sim -1.5$. 
This partially accounts for the 
difference between average values in \citet{kov01} and in our 
distribution in Fig. 4, but there still is a residual difference of 
0.17 
dex between 
the average values,   
 and our distribution
appears to cover a metallicity range larger than in \citet{kov01}.

We determined metallicities for six of the RRd variables discovered by 
\citet{kov01}, three of which are actually only suspected RRd's.
From our re-analysis of the light curves 
we think that these stars are monoperiodic {c-}type RR Lyraes with noisy light curves.
The comparison between individual metallicity values is shown in Table~\ref{t:tabrrd},
where the range given in Column 4 was evaluated from the metallicity 
distribution of the RRd stars in 
fig. 5 of \citet{kov01}. Unfortunately we did not observe the two
most metal poor RRd's in \citet{kov01}. 

\begin{table*}
\caption{Comparison of the metallicities for RRd variable stars in Sculptor}
\begin{tabular}{rccccc}
\hline
Star&[Fe/H]$_{\rm this-paper}$&[Fe/H]$_{\rm this-paper}$&[Fe/H]& Type&Type  \\
    &  ZW84                   &   CG97                  & \citet{kov01} & \citet{kov01}& this paper\\  
\hline
   59&$-$2.14&$-$1.97&$-1.6\leq$[Fe/H]$\leq -1.4$& RRd&RRc\\
 1439&$-$1.88&$-$1.70&$-1.6\leq$[Fe/H]$\leq -1.4$& RRd&RRc\\
 3016&$-$1.81&$-$1.61&$-1.6\leq$[Fe/H]$\leq -1.4$& RRd&RRc\\
 3044&$-$1.86&$-$1.67&$-1.6\leq$[Fe/H]$\leq -1.4$& RRd&RRd\\
 3104&$-$1.51&$-$1.30&$-1.6\leq$[Fe/H]$\leq -1.4$& RRd&RRd\\
 3143&$-$1.36&$-$1.14&$-1.6\leq$[Fe/H]$\leq -1.4$& RRd&RRd\\
\hline
\end{tabular}
\label{t:tabrrd}
\end{table*}
As with the RRab's the metallicity
range spanned by the RRd stars we analyzed is  
larger than that obtained by 
\citet{kov01} for the same stars.

\section{The luminosity-metallicity relation}

The luminosity-metallicity relation followed by the RR Lyrae stars in
Sculptor was derived 
using 
the intensity-averaged mean
magnitudes 
and the metal abundances
in Table~\ref{t:sculptor}.
We first discarded all objects that are not RR Lyrae stars or have incomplete
light curves (stars 3302, 3710, 4780, and 5724). We also
eliminated the most metal rich variable in our sample (star 4263), even if 
this star falls extremely well on the
mean relations we derive. 
Following the procedure applied by \citet{g04} to 
the LMC RR Lyrae stars, we divided the Sculptor 
variables into 6 metallicity bins 0.1 dex wide; the 
corresponding
average apparent magnitudes are given in Table~\ref{t:tabbin}
in the two
metallicity scales, respectively. A least square fit 
weighted by the
errors in both variables gives: 

$$\langle V\rangle = (0.092 \pm 0.027)({\rm [Fe/H]}_{\rm ZW}+1.5)+(20.158 \pm 0.009)$$

\noindent
where the errors in the slopes were evaluated via Monte Carlo simulations. 
The same slope is found for metallicities in CG97 scale.

\begin{table*}
\caption{Metallicity distribution and average magnitudes of the RR Lyrae
stars in Sculptor}
\begin{tabular}{lcccc}
\hline
$[$Fe/H$]_{ZW}$~ bin & n. stars & $\langle V\rangle$ & $\langle [$Fe/H]$_{\rm ZW}\rangle$ & $\langle [$Fe/H]$_{\rm CG}\rangle$  \\
\hline
~~~~~$[$Fe/H$] \leq -2.0$ & 22 & 20.093$\pm$0.017 & $-$2.175 $\pm$ 0.034 & $-$2.005 $\pm$0.036 \\
$-2.0<[$Fe/H$]<-1.9$      & 14 & 20.139$\pm$0.017 & $-$1.957 $\pm$ 0.008 & $-$1.773 $\pm$0.008 \\
$-1.9<[$Fe/H$]<-1.8$      & 19 & 20.099$\pm$0.016 & $-$1.846 $\pm$ 0.006 & $-$1.656 $\pm$0.007 \\
$-1.8<[$Fe/H$]<-1.7$      & 23 & 20.140$\pm$0.015 & $-$1.757 $\pm$ 0.005 & $-$1.559 $\pm$0.006 \\
$-1.7<[$Fe/H$] \leq -1.6$ & 13 & 20.144$\pm$0.018 & $-$1.638 $\pm$ 0.009 & $-$1.432 $\pm$0.010 \\
$-1.6<[$Fe/H$]$      & 14 & 20.163$\pm$0.012 & $-$1.461 $\pm$ 0.018 & $-$1.245 $\pm$0.019 \\
\hline
\end{tabular}
\label{t:tabbin}
\end{table*}

\begin{figure} 
\includegraphics[width=8.8cm, bb=35 385 560 675]{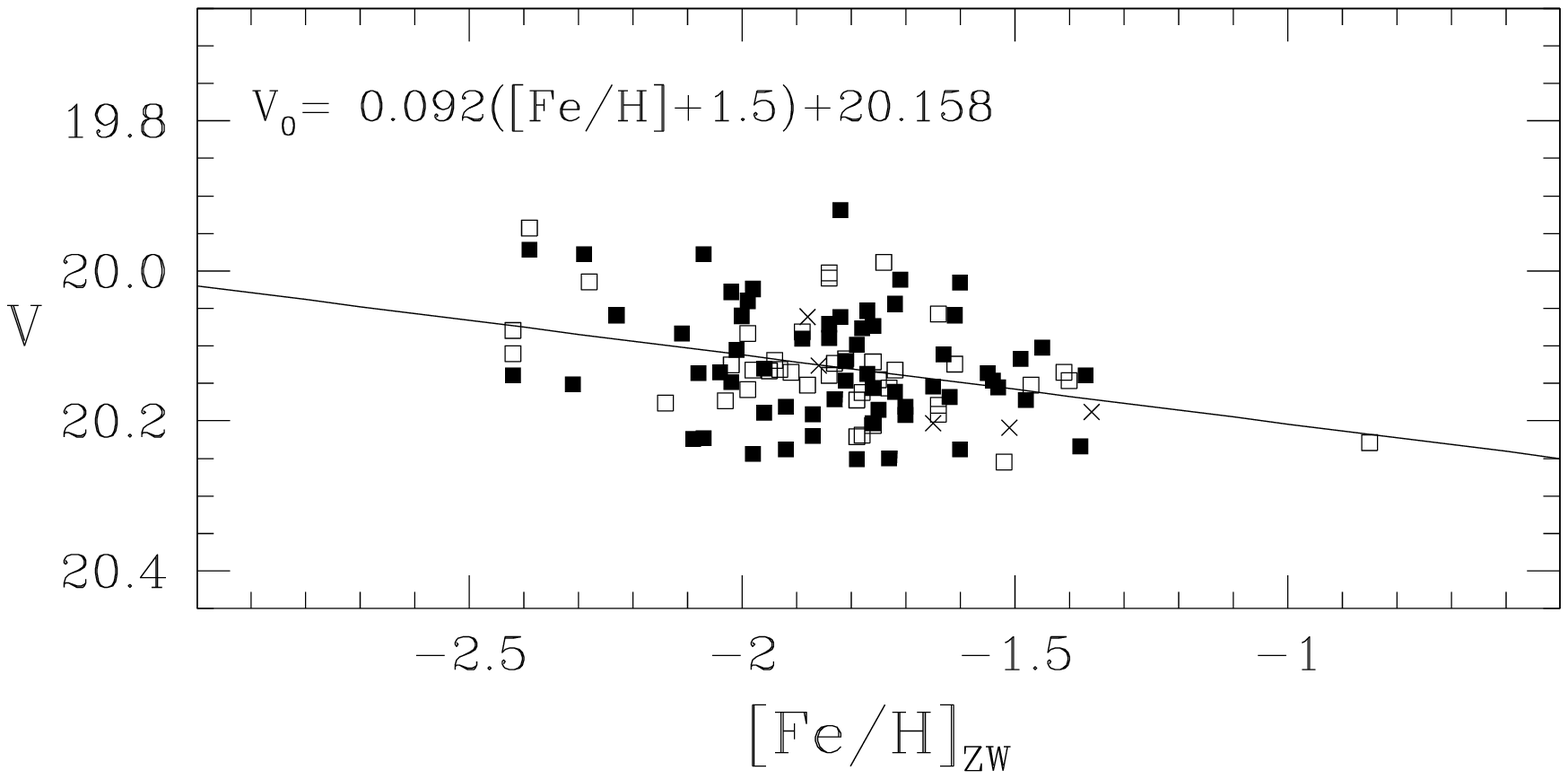}
\caption[]{Luminosity-metallicity relation of the RR Lyrae stars in 
Sculptor; variables are divided by type, filled squares: RRab pulsators,
open squares: RRc pulsators, crosses: RRd pulsators.
}
\label{f:fig3t}
\end{figure}

The luminosity-metallicity relation of the Sculptor RR Lyrae stars is 
shown in 
Fig.~\ref{f:fig3t}.
It is based on 105 stars covering the metallicity range [Fe/H]
from
$-1.36$ to $-2.42$, and we 
used different symbols for the various types of
RR Lyrae stars. All variables seem to follow the same luminosity-metallicity
relation independent of type.
\citet{g04} found that the LMC 
double-mode RR Lyrae stars are offset to brighter luminosities
in the luminosity-metallicity plane and explain 
this evidence with the LMC RRd's being more evolved than the single-mode
pulsators. 
The lack of a similar difference in luminosity between single
and double-mode Sculptor RR Lyrae stars suggests that in this galaxy 
also the single-mode variables
are evolved. 

A more striking difference 
is 
 the slope 
of the 
luminosity-metallicity 
relation,
significantly shallower than that 
obtained 
for the LMC variables
by \citet{g04}\footnote{
However, we note that in both studies the slopes of the RR Lyrae
luminosity-metallicity relations could be 
slightly underestimated by a few hundredths of magnitude, because of the
relatively large errors of the individual [Fe/H] values.}.
On the 
other
hand, we note that also the scatter in the average apparent luminosity of the RR 
Lyrae stars
in Sculptor is less than half of that observed for the variables in the LMC: 
$\sigma _V$(Sculptor)=0.07 mag, compared to the 0.15-0.16 mag in the
LMC (\citealt{clem03b}). This scatter is almost entirely accounted for by 
photometric errors and the dispersion in metallicity of the Sculptor
RR Lyrae stars, thus indicating that these variables
have a very similar degree of evolution off the zero age horizontal
branch. The shallow slope of the luminosity-metallicity relation
in Sculptor could thus be explained with the Sculptor RR Lyrae's
being  all rather evolved stars 
arising from the old metal poor population 
that also gives origin to the blue HB in the
galaxy. 
Some support to this suggestion arises from the comparison of  the
average luminosity of the RR Lyrae stars with that of non variable 
horizontal branch stars in Sculptor.

From the color-magnitude diagram of Sculptor, values of 
$\langle V_{\rm HB}\rangle \simeq$ 20.18-20.20 mag, 
$\langle V_{\rm blue~HB} \rangle \simeq $ 20.20 and $\langle
 V_{\rm red~HB} \rangle \simeq$ 20.29 mag are obtained by \citet{kal95},
  Babusiaux, Gilmore, \& Irwin (2005), and \citet{rizphd}, respectively.
This is at least $\simeq$0.06-0.08 mag
fainter than the average luminosity of the RR Lyrae stars
which, for the 106 RR Lyrae with full light curve coverage, 
is $\langle V\rangle$=20.127 with a dispersion of $\sigma$=0.072 (this value
agrees well with the estimate by \citealt{kal95}, based on the total sample
of 226 RR Lyrae stars in Sculptor). 

\section{Radial velocity determinations}

Radial velocities were measured from the spectra of the Sculptor 
RR Lyrae stars, and are given in Column 16 of 
Table~\ref{t:sculptor}.
 Multiple observations of the same stars show that our estimates
have typical errors of about $\pm 15$ km s$^{-1}$, with no systematic 
differences 
for different masks.  This error includes the contribution
of measurement uncertainties, errors related to the centering of the stars
in the slit, and uncertainties due to the poor sampling of the radial velocity
curves of the variable stars.
%

The average heliocentric radial velocity of all our variable stars in
Sculptor is 
$\langle V_r\rangle$=$109.1\pm$ 1.9\,km s$^{-1}$ (r.m.s.=19.9 km s$^{-1}$,
110 stars, having preliminarily averaged individual values for 
stars with multiple observations). The average radial velocity of the RR Lyrae 
stars alone is
$\langle V_r\rangle$=$109.6 \pm 1.9 $ km s$^{-1}$
 (r.m.s.=19.8 km s$^{-1}$,
107 stars).
These values are 
in excellent agreement with the estimate 
obtained from K-giants in Sculptor by 
Queloz, Dubath \& Pasquini (1995: 
$\langle V_r\rangle$=$109.9 \pm 1.4 $ km s$^{-1}$) 
 and are consistent with the 
value of  $\langle V_r\rangle$=$107 \pm 2.0$ km s$^{-1}$
previously derived
by Armandroff \& Da Costa (1986).

 The good agreement with the literature
values suggests that the offcentering problems noted for the calibrating 
cluster variables (see Appendix B) do not seem to affect the variables in Sculptor
and that the undersampling of the variable star's radial velocity
curves does not significantly bias our estimates of the average $V_r$ value. 
 To further check this point we extracted from the 
 database of the Galactic field RR Lyrae stars analyzed with the 
 Baade-Wesselink 
 method (\citealt{liu90,jones92,cacciari92,skillen93,fernley94}) template radial velocity
 curves of {\it ab-} and {\it c-}type RR Lyrae stars with metal
 abundance comparable to that of the variable stars in Sculptor, and estimated 
 phase-dependent radial velocity corrections for each spectrum of 
 RR Lyrae star. Using this procedure we found that the average 
 correction to apply to the radial velocity 
 mean value of the
 107 RR Lyrae stars in our sample 
 is less than $\sim$0.1 km s$^{-1}$ and
 can be safely neglected.  
 
The difference between r.m.s. scatter of the RR Lyrae stars 
(19.8 km s$^{-1}$) and typical
measurement errors (15 km s$^{-1}$) implies an intrinsic radial velocity dispersion 
of $\sigma$=12.9 km s$^{-1}$ for the RR Lyrae stars. This value is larger
than the $6.3+1.1,-1.3$ km s$^{-1}$ and $6.2\pm 1.1$ km s$^{-1}$  found for the 
K-giants by Armandroff \& Da Costa (1986) and \citet{queloz95},  
and for Sculptor metal rich red giant stars 
by Tolstoy et al. 
(2004; $\sigma_{metal-rich}$=$7\pm 1$ km s$^{-1}$), but is consistent with
the  $11\pm 1$ km s$^{-1}$ dispersion observed by these same authors for the
metal poor red giants
in Sculptor. 
This result gives further support
to the hypothesis that the RR Lyrae stars in Sculptor arise from the old, 
metal-poor population giving origin to the galaxy blue-horizontal branch, 
although our value for their velocity dispersion
 needs to be confirmed by higher resolution spectroscopy.

\section{Metallicity distributions of the different old
stellar components in Sculptor}

The average metallicity, metal abundance distribution and range in metal abundance
spanned by the RR Lyrae stars can be compared with the analogous quantities
for other old stellar components in Sculptor, namely with the 
metallicity spread inferred from the width of the red giant branch 
(e.g. \citealt{kal95, maj99, rizphd, babusiaux05}),
and 
with the abundances 
directly measured for red giants
in Sculptor by  \citet{gei05} and 
\citet{tol04}, respectively. 
 
The metallicity distribution in Fig.~\ref{f:fig3hbw}
shows that the RR Lyrae stars in Sculptor cover a full metallicity range 
of about 1.6 dex, which however reduces to $\sim$ 1 dex if the 
single most metal rich star in the sample is discarded.
This range is larger than inferred from the spread 
of the red giant 
stars ($\Delta {\rm [Fe/H]} \simeq$ 0.6 dex, \citealt{kal95, rizphd}; 
$\simeq$0.8 dex, \citealt{maj99}; $\simeq$0.7 dex, \citealt{babusiaux05}), 
and is consistent with
the spectroscopic study of red giants in Sculptor by
Geisler et al. (2005:  $\simeq$1.1 dex) and 
\citet{tol04} in the galaxy inner region (see upper panel of their fig. 3). 

\citet{tol04} find that 
the ancient stellar component ($\geq$ 10 Gyr old)  
in Sculptor is divided into two distinct groups
having different metal abundance, kinematics and spatial distribution.
The metal-rich population is concentrated within the $r = 0.2$  degree 
central region,  
has metallicities in the range $-1.7<$[Fe/H]$<-0.9$, velocity dispersion of 
$\sigma_{metal-rich}$=7$\pm$1 km s$^{-1}$ and is related to the 
Sculptor red horizontal branch.
The metal-poor population is more spatially extended,
has metallicities in the range $-2.8<$[Fe/H]$<-1.7$, velocity dispersion of 
$\sigma_{metal-poor}$=11$\pm$1 km s$^{-1}$ and is related to the 
Sculptor blue horizontal branch.

Our RR Lyrae stars are located in the central region of Sculptor,
virtually coincident with the region where \citet{tol04} find
segregation of red HB stars. However, their metallicity distribution
is dominated by the metal-poor objects with an average value of
[Fe/H]=$-$1.83.
The 26 stars with [Fe/H]$>-$1.7 are  
marked by boxes in Fig.~\ref{f:fatte}.
Their average luminosity $\langle V_{\rm [Fe/H]>-1.7}\rangle$=20.154 
($\sigma$=0.060,
26 stars) is marginally fainter that the average of the
remaining 80 stars with  [Fe/H]$\leq -$1.7 
($\langle V_{\rm [Fe/H]\leq -1.7}\rangle$=20.118, $\sigma$=0.075,
80 stars), as expected given the shallow slope of the 
Sculptor RR Lyrae luminosity-metallicity relation, and is anyway  
brighter than both the blue and red HBs of the non variable stars, thus 
indicating that  they are evolved 
objects, like their metal-poor counterpart. 
The average radial velocity of the two samples is only marginally different: 
$\langle v_r\rangle$=117.5 (r.m.s.=18.5, 26 stars) for the metal-rich
sample and $\langle v_r\rangle$=107.0 (r.m.s.=19.6, 80 stars) for the 
metal-poor stars. The r.m.s. scatters are very similar and, once deconvolved
for the measurement errors (15 km s$^{-1}$), lead to very similar velocity dispersions
of 10.8 and 12.6 km s$^{-1}$ in agreement with the velocity dispersion
measured by \citet{tol04} for the metal-poor component associated to 
Sculptor blue HB.

\section{Summary and conclusions}

Low resolution spectra obtained with FORS2 at the VLT have been 
used to measure individual metal abundances [Fe/H] and radial velocities 
for 107 RR Lyrae stars in the Sculptor dwarf spheroidal galaxy.
Metallicities were derived using a revised version of the
$\Delta$S method \citep{g04}. The RR Lyrae stars in Sculptor are predominantly
metal-poor with an average metal abundance of [Fe/H]=$-1.83 \pm 0.03$ 
(r.m.s.=0.26 dex) on the ZW84 metallicity scale, and only a few outliers 
having metallicities larger than $-1.4$ dex.
The observed metallicity dispersion is larger than the observational errors,
thus showing that these variables 
have a real metallicity spread.

The RR Lyrae stars in Sculptor are found to follow a 
luminosity-metallicity relation with a 
slope of 
0.09 mag dex$^{-1}$, which is shallower than in the LMC \citep{g04}.
This is explained with the Sculptor variable stars being rather evolved from
the zero age HB, as also supported by their brighter luminosity compared to 
the non variable HB stars.

From our spectra we measured an intrinsic velocity dispersion 
12.9 km s$^{-1}$ for the RR Lyrae stars, which appears to be 
 in agreement with the dispersion derived by 
\citet{tol04} for metal-poor red giants associated to the blue-HB stars in
Sculptor. 

All these evidences suggest that our RR Lyrae sample, and the 
RR Lyrae stars in Sculptor in general, 
are connected to the blue-HB population
and arise 
from the first burst of star formation that produced the galaxy blue metal-poor HB. 
They allow to trace and distinguish
this older component in the internal regions of the galaxy which are otherwise 
dominated by
the metal-rich and younger red-HB stars.
Indeed, this component only produced a few, if any, of the Sculptor 
RR Lyrae stars
since average luminosity and kinematic properties of the few 
metal-rich objects 
in our sample suggest that they are more likely the high metallicity tail of the metal-poor star
population rather than objects associated with Sculptor  red-HB stars.

\section*{Acknowledgments}
A special thanks goes to J. Kaluzny for providing us 
the time series photometry of the Sculptor variables.
We thank the anonymous referee for his/her comments and suggestions.
This research was funded 
by MIUR, under the scientific projects: 2002028935, ``Stellar Populations in the Local Group'' 
(P.I.: Monica Tosi) and  2003029437, ``Continuity and Discontinuity in the 
Milky Way Formation'' (P.I.: Raffaele Gratton).

\section{Appendix A - Finding charts}
We present in this Section finding charts (Fig.s 6a,b,c) for all 
the variable stars we observed in Sculptor. 
They correspond to the nine  $6.8\arcmin \times 6.8 \arcmin$ 
FORS2 subfields used to map the central 
 $15\arcmin \times 15 \arcmin$ area of the galaxy.
Centre of field coordinates are provided in Table~\ref{t:tablelog}.  
In each map North is up and East to the left, and
variables are identified according to \citet{kal95} 
identifiers.   
The RR Lyrae stars are marked by open circles (in 
red in the electronic edition of the journal), other types
of variables by (blue) open squares.
The complete listing of the variables observed in Sculptor is provided in 
Table~\ref{t:sculptor}. Equatorial
coordinates for all our targets can be found in table~2 of \citet{kal95}.






\section{Appendix B - The metallicity calibration}

Following the procedure devised by \citet{g04}, 
line indices for RR Lyrae stars are computed from the spectra shifted to rest
wavelength by directly integrating the instrumental fluxes in spectral bands 
centered on the Ca II K, H$\delta$, H$\gamma$, and H$\beta$ lines 
(see table~2 and fig. 10 of 
\citealt{g04} for the definition of the spectral bands). Then a 
$\langle H\rangle$ index is defined 
as the average of the
indices of the 3 hydrogen lines, and  $K$  as the index of the Ca II
K line. Metallicities are derived by comparing 
the $\langle H\rangle$ and $K$  indices measured for the target stars 
to the same
quantities for variables in a number of globular clusters of 
known metal abundance.


The calibration of the line indices of the Sculptor variables 
in terms of metal abundances [Fe/H] was obtained using RR Lyrae stars 
in the newly observed clusters M\,15, M\,2 and NGC\,6171, and in the 
calibrating 
clusters used in \citet{g04}, namely M\,68, NGC\,1851 and NGC\,3201.
For all these clusters, precise metal abundances are available on both 
the ZW84 and CG97  
metallicity scales. 
NGC\,6441, a metal rich cluster ([Fe/H]=$-0.59$ according to ZW84
metallicity scale) having  RR Lyrae stars
with anomalously long periods, was not used for calibration purposes 
and will be discussed
elsewhere (see \citealt{clem05}).

Line indices measured for RR Lyrae stars in the calibrating 
clusters analyzed in the present paper are provided in 
Tables~\ref{t:M15},~\ref{t:M2}, and 
\ref{t:NGC6171} for M\,15, M\,2 and NGC\,6171, respectively.
\begin{figure} 
\includegraphics[width=7cm, bb=140 140 440 720]{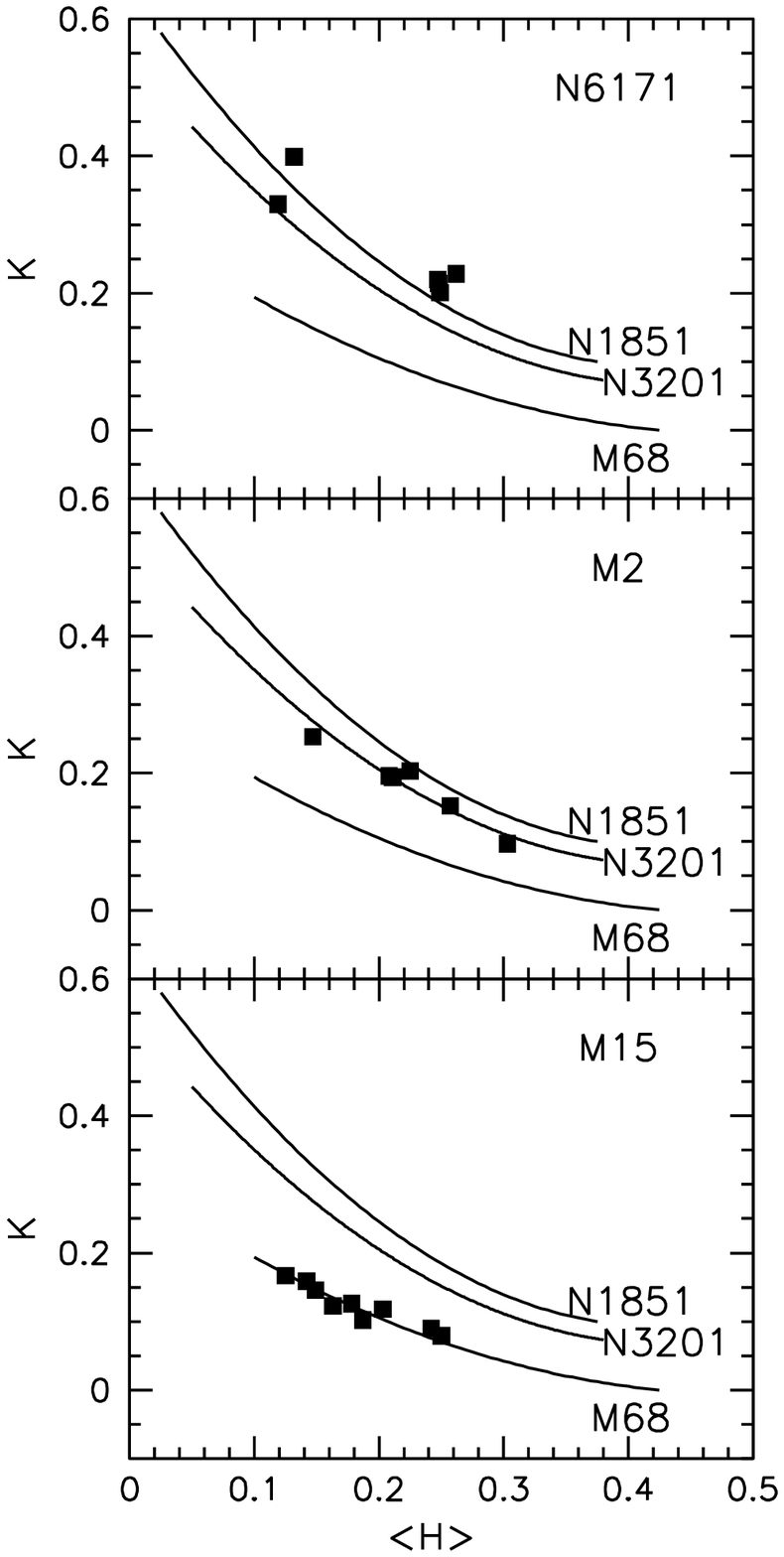}
{\medskip
\par\noindent
Figure 7. Correlation between $K$ and $\langle H\rangle$ spectral indices for the
calibrating clusters M\,15, M\,2, and NGC\,6171 observed in the 
present run. Solid lines represent the ~~mean lines for the calibrating clusters
M\,68, NGC\,3201, and NGC\,1851, from \citet{g04}.}
\label{f:cal01}
\end{figure}
The calibrating clusters define mono-metallic  
correlations in the $K$ {\it versus} $\langle H\rangle$ plane. These relations   
 are shown in Fig.~7
 where filled squares represent stars
in  M\,15, M\,2 and NGC\,6171, 
and solid lines are the mean relations 
defined by M\,68, NGC\,1851 and NGC\,3201 taken from \citet{g04}.
RR Lyrae stars in M\,15 fall precisely on the mean line of M\,68 and those in 
M\,2 closely follow the mean line of NGC\,3201,  
confirming the good agreement between metal abundances 
of these two pairs
of clusters. Stars in NGC\,6171 generally
fall slightly above the mean line of NGC\,1851, in agreement with the slightly higher
metallicity of NGC\,6171 with respect to NGC\,1851. 

According to \citet{g04} the mean relations drawn in 
Fig.~
7
by  M\,68 and NGC\,1851 are:
\begin{equation}
K_1 = 0.3093 - 1.2815~\langle H\rangle + 1.3045~\langle H^2\rangle
\end{equation}
(valid for $\langle H\rangle$\ between 0.12 and 0.40) for M~68, and:
\begin{equation}
K_2 = 0.6432 - 2.6043~\langle H\rangle + 3.0820~\langle H^2\rangle
\end{equation}
(valid for $\langle H\rangle$\ between 0.04 and 0.34) for NGC 1851.
We thus  
defined metallicity index $M.I.$ the quantity:
\begin{equation}
M.I.= (K - K_1)/(K_2 - K_1)
\end{equation}
where $K$ is the Ca II K line index of the star, and $K_1, K_2$ are derived
entering  the $\langle H\rangle$ index measured for the star into equations (1) and (2). 
$M.I.$ values derived by this procedure for the RR Lyrae stars in the
 calibrating 
clusters analyzed in the present paper are listed in 
Column 10   
Tables~\ref{t:M15}, ~\ref{t:M2}, ~\ref{t:NGC6171}. 

The calibration of the metallicity index in terms of metal abundance
[Fe/H] was derived by computing average $\langle M.I.\rangle$ values 
for the M\,15, M\,2 and NGC\,6171 variables 
from the individual $M.I.$'s in 
Tables~\ref{t:M15}, ~\ref{t:M2}, ~\ref{t:NGC6171} 
and from the $M.I.$ values in
Column 9 of 
tables 3, 4, and 5 of \citet{g04} for M\,68, NGC\,1851 and NGC\,3201, 
and correlating these $\langle M.I.\rangle$'s with the
metal abundances of the clusters on the ZW and CG metallicity scales,
respectively. The average $\langle M.I.\rangle$ values and their dispersions are summarized
in Columns 6 and 7 of Table~\ref{t:calibration} 
along with the cluster metallicities (and their uncertainties) in the two 
metallicity scales
(Columns from 2 to 5).

The correlation between [Fe/H] and $\langle M.I.\rangle$ values is very 
well represented by linear
regressions with scatter typically within the error of measure.
These linear regressions are described by
the following equations:
\begin{equation}
{\rm [Fe/H]_{ZW}}= 0.882 \langle M.I.\rangle - 2.170
\end{equation}
\begin{equation}
{\rm [Fe/H]_{CG}}= 0.941 \langle M.I.\rangle - 2.000 
\end{equation} 

   \begin{table*}
      \caption[]{Line indices and metal abundances of RR Lyrae stars 
      in the globular cluster M~15}
         \label{t:M15}
     $$
         \begin{array}{cccccccccrccrp{0.5\linewidth}cccccccccccc}
            \hline
            \noalign{\smallskip}
            \multicolumn{12}{c}{\rm M\,15} \\
            \hline
            \noalign{\smallskip}
            {\rm Star} & {\rm HJD} & {\rm Type} & \phi &  K & {\rm H}_{\delta} & {\rm H}_{\gamma} & 
	    {\rm H}_{\beta} & \langle H\rangle  & M.I. & {\rm [Fe/H]}   &   {\rm [Fe/H]} &  V_{r}~~\\
            {\rm (a)}& {\rm (-2400000)} &  & {\rm (b)} &   &           &            &            &     
	    &      &   {\rm ZW}   &     {\rm CG}  & {\rm km s^{-1}}\\
            \noalign{\smallskip}
            \hline
            \noalign{\smallskip}
            {\rm V10}& 52851.6744 &{\rm c} & 0.781 &  0.126  & 0.195 & 0.192 & 0.147 & 0.178 &  0.025 & -2.15 &  -1.98 & -83~~\\
            {\rm V12}& 52851.6744 &{\rm ab}& 0.645 &  0.159  & 0.148 & 0.155 & 0.122 & 0.142 &  0.027 & -2.15 &  -1.98 & -92~~\\
            {\rm V17}& 52851.6744 &{\rm d }& 0.429 &  0.123  & 0.181 & 0.151 & 0.158 & 0.163 & -0.073 & -2.23 &  -2.07 & -96~~\\
            {\rm V19}& 52851.6744 &{\rm ab}& 0.592 &  0.090  & 0.272 & 0.247 & 0.208 & 0.242 &  0.121 & -2.06 &  -1.89 &-122~~\\
            {\rm V23}& 52851.6744 &{\rm ab}& 0.024 &  0.118  & 0.218 & 0.208 & 0.184 & 0.203 &  0.108 & -2.08 &  -1.90 &-118~~\\
            {\rm V25}& 52851.6744 &{\rm ab}& 0.492 &  0.167  & 0.141 & 0.107 & 0.126 & 0.125 & -0.015 & -2.18 &  -2.01 & -99~~\\
            {\rm V29}& 52851.6744 &{\rm ab}& 0.967 &  0.146  & 0.170 & 0.145 & 0.133 & 0.149 & -0.002 & -2.17 &  -2.00 &-114~~\\
            {\rm V42}& 52851.6744 &{\rm c }& 0.818 &  0.102  & 0.228 & 0.171 & 0.163 & 0.187 & -0.092 & -2.25 &  -2.09 & -99~~\\
            {\rm V50}& 52851.6744 &{\rm c }& 0.298 &  0.079  & 0.272 & 0.253 & 0.225 & 0.250 &  0.078 & -2.10 &  -1.93 &-104~~\\
            \noalign{\smallskip}	  
            \hline
         \end{array}
     $$
\begin{list}{}{}
\item[$^{\mathrm{a}}$] Star identifiers are from Sawyer-Hogg on line catalogue of variable stars in Galactic
globular clusters published by \citet{clement01}
\item[$^{\mathrm{b}}$] Phases were calculated from the HJD of 
observation using epochs and periods corrected according to 
the rates of period changes published by \citet{sil95}. 
\end{list}
   \end{table*}

   \begin{table*}
      \caption[]{Line indices and metal abundances of RR Lyrae stars 
      in the globular cluster M~2}
         \label{t:M2}
     $$
         \begin{array}{lccccccccccccp{0.5\linewidth}cccccccccccc}
            \hline
            \noalign{\smallskip}
            \multicolumn{12}{c}{\rm M2} \\
            \hline
            \noalign{\smallskip}
            {\rm Star} & {\rm HJD} & {\rm Type} & \phi &  K & {\rm H}_{\delta} & {\rm H}_{\gamma} & 
	    {\rm H}_{\beta} & \langle H\rangle  & M.I. & {\rm [Fe/H]}   &   {\rm [Fe/H]} &  V_{r}~~\\
            {\rm (a)}& {\rm (-2400000)} &  & {\rm (b)} &   &           &            &            &     
	    &      &   {\rm ZW}   &     {\rm CG}  & {\rm km s^{-1}}\\
	    \hline
	    \noalign{\smallskip}
 {\rm V10  }& 52856.7503 &{\rm ab }& 0.758 & 0.194 & 0.221 & 0.219 & 0.193 & 0.211 & 0.722 & -1.53 & -1.32 & -38~\\
 {\rm V15  }& 52856.7503 &{\rm c/d}& 0.345 & 0.097 & 0.319 & 0.308 & 0.281 & 0.303 & 0.583 & -1.66 & -1.45 & -25~\\
 {\rm V26^c}& 52856.7503 &{\rm c  }& 0.428 & 0.196 & 0.224 & 0.223 & 0.177 & 0.208 & 0.716 & -1.54 & -1.33 & -21~\\
 {\rm V32^c}& 52856.7503 &{\rm c  }& 0.350 & 0.152 & 0.266 & 0.257 & 0.249 & 0.257 & 0.774 & -1.49 & -1.27 & -34~\\
 {\rm V7   }& 52856.7503 &{\rm ab }& 0.621 & 0.203 & 0.221 & 0.217 & 0.238 & 0.225 & 0.923 & -1.36 & -1.13 & -36~\\
 {\rm V3^d }& 52856.7503 &{\rm ab }& 0.191 & 0.253 & 0.150 & 0.144 & 0.147 & 0.147 & 0.585 & -1.65 & -1.45 & -15~\\
            \noalign{\smallskip}
            \hline
         \end{array}
     $$
\begin{list}{}{}

\item[$^{\mathrm{a}}$] Star identifiers are from Sawyer-Hogg on line catalogue of variable stars in Galactic
globular clusters published by \citet{clement01}
\item[$^{\mathrm{b}}$] Phases of the spectra were derived from the HJD of observation and the 
ephemerides published by  
\citet{lc99}.
\item[$^{\mathrm{c}}$] Variable stars V26 and V32 correspond to stars LC608 and
LC864 of \citet{lc99}, respectively.
\item[$^{\mathrm{d}}$] Doubtful identification.
\end{list}
   \end{table*}

   \begin{table*}
      \caption[]{Line indices and metal abundances of RR Lyrae stars 
      in the globular cluster NGC~6171}
         \label{t:NGC6171}
     $$
         \begin{array}{lccccccccccccp{0.5\linewidth}cccccccccccc}
            \hline
            \noalign{\smallskip}
            \multicolumn{12}{c}{\rm NGC\,6171} \\
            \hline
            \noalign{\smallskip}
            {\rm Star} & {\rm HJD} & {\rm Type} & \phi &  K & {\rm H}_{\delta} & {\rm H}_{\gamma} & 
	    {\rm H}_{\beta} & \langle H\rangle  & M.I. & {\rm [Fe/H]}   &   {\rm [Fe/H]} &  V_{r}~~\\
            {\rm (a)}& {\rm (-2400000)} &  & {\rm (b)} &   &           &            &            &     
	    &      &   {\rm ZW}   &     {\rm CG}  & {\rm km s^{-1}}\\
            \noalign{\smallskip}
            \hline
            \noalign{\smallskip}
  {\rm V2 }& 52849.5952 &{\rm ab}& 0.746 & 0.330 & 0.113 & 0.128 & 0.117 & 0.119 & 0.771 & -1.49 & -1.27 & -77~\\
  {\rm V4 }& 52849.5952 &{\rm c }& 0.385 & 0.220 & 0.283 & 0.241 & 0.219 & 0.247 & 1.280 & -1.04 & -0.80 & -91~\\
  {\rm V6 }& 52849.5952 &{\rm c }& 0.570 & 0.214 & 0.256 & 0.284 & 0.205 & 0.248 & 1.239 & -1.08 & -0.83 & -87~\\
  {\rm V7 }& 52849.5952 &{\rm ab}& 0.683 & 0.399 & 0.124 & 0.129 & 0.142 & 0.132 & 1.234 & -1.08 & -0.84 & -86~\\
  {\rm V9 }& 52849.5952 &{\rm c }& 0.793 & 0.201 & 0.277 & 0.238 & 0.231 & 0.249 & 1.133 & -1.17 & -0.93 & -96~\\
  {\rm V21}& 52849.5952 &{\rm c }& 0.774 & 0.228 & 0.267 & 0.295 & 0.222 & 0.262 & 1.504 & -0.84 & -0.59 & -77~\\
            \noalign{\smallskip}
            \hline
         \end{array}
     $$
\begin{list}{}{}
\item[$^{\mathrm{a}}$] Star identifiers are from Sawyer-Hogg on line catalogue of variable stars in Galactic
globular clusters published by \citet{clement01}
\item[$^{\mathrm{b}}$] Phases were derived from the HJD of observation, the periods published by \citet{cs97}  and the epochs we
estimated from the study of the light curves published from the same authors. V2 was not observed by 
\citet{cs97}, for this star the adopted ephemerides, from \citet{csh71}, may be no longer valid. 
\end{list}
   \end{table*}

   \begin{table*}
      \caption[]{Calibration of the metallicity index}
	 \label{t:calibration}
     $$
	 \begin{array}{lccccccccccrp{0.5\linewidth}cccccccc}
	    \hline
	    \noalign{\smallskip}
	    {\rm Cluster} & {\rm [Fe/H]} & {\rm err} & {\rm [Fe/H]} & {\rm err} &  \langle M.I.\rangle &
	    {\rm err} & \langle {\rm [Fe/H]}\rangle   & \sigma& \langle {\rm [Fe/H]}\rangle  & \sigma&
	    \langle V_{r}\rangle ~~ \\
		    &	{\rm ZW}   &     &   {\rm CG}   &	  &	  &	&  {\rm This~paper}&    
		      & {\rm This~paper}&	 & {\rm km~s^{-1}}\\   
		    &	     &     &	    &	  &	  &	&  {\rm ZW~scale}  &       & {\rm
		    CG~scale}  &	 &	   \\	
	    \noalign{\smallskip}
	    \hline
	    \noalign{\smallskip}
	   {\rm NGC\,1851}	 & -1.36 & 0.09 & -1.03 & 0.06 & 1.007 & 0.028 &-1.36\pm 0.02^{\mathrm{a}}&0.10&~-&-&-~~\\
	   {\rm NGC\,3201}	 & -1.61 & 0.12 & -1.24 & 0.03 & 0.710 & 0.030 &-1.56\pm 0.02^{\mathrm{a}}&0.08&~-&-&-~~\\
	   {\rm NGC\,4590~(M68)} & -2.09 & 0.11 & -2.00 & 0.03 & 0.000 & 0.014 &-2.09\pm 0.01^{\mathrm{a}}&0.05&~-&-&-~~\\
	   {\rm NGC\,6171}	 & -0.99 & 0.06 & -0.95 & 0.04 & 1.194 & 0.098 &-1.12\pm 0.09~&0.21&-0.88\pm 0.09&0.22& -86 \pm 3\\
	   {\rm NGC\,7078~(M15)} & -2.15 & 0.08 & -2.02 & 0.04 & 0.020 & 0.025 &-2.15\pm 0.02~&0.06&-1.98\pm 0.02&0.07&-103 \pm 4\\
	   {\rm NGC\,7089~(M2)}  & -1.62 & 0.07 & -1.31 & 0.04 & 0.717 & 0.052 &-1.54\pm 0.05~&0.11&-1.33\pm 0.05&0.12& -28 \pm 4\\
	    \noalign{\smallskip}
	    \hline
	 \end{array}
     $$
\begin{list}{}{}
\item[$^{\mathrm{a}}$] Averages of the values in tables 3,4, and 5 of \citet{g04} 
corrected downward respectively by
0.10, 0.07 and 0.03 dex to account for systematic differences between ZW84 and the metallicity scale adopted
in \citet{g04}.  
\end{list}
   \end{table*}

\begin{figure*}
\includegraphics[width=8.8cm]{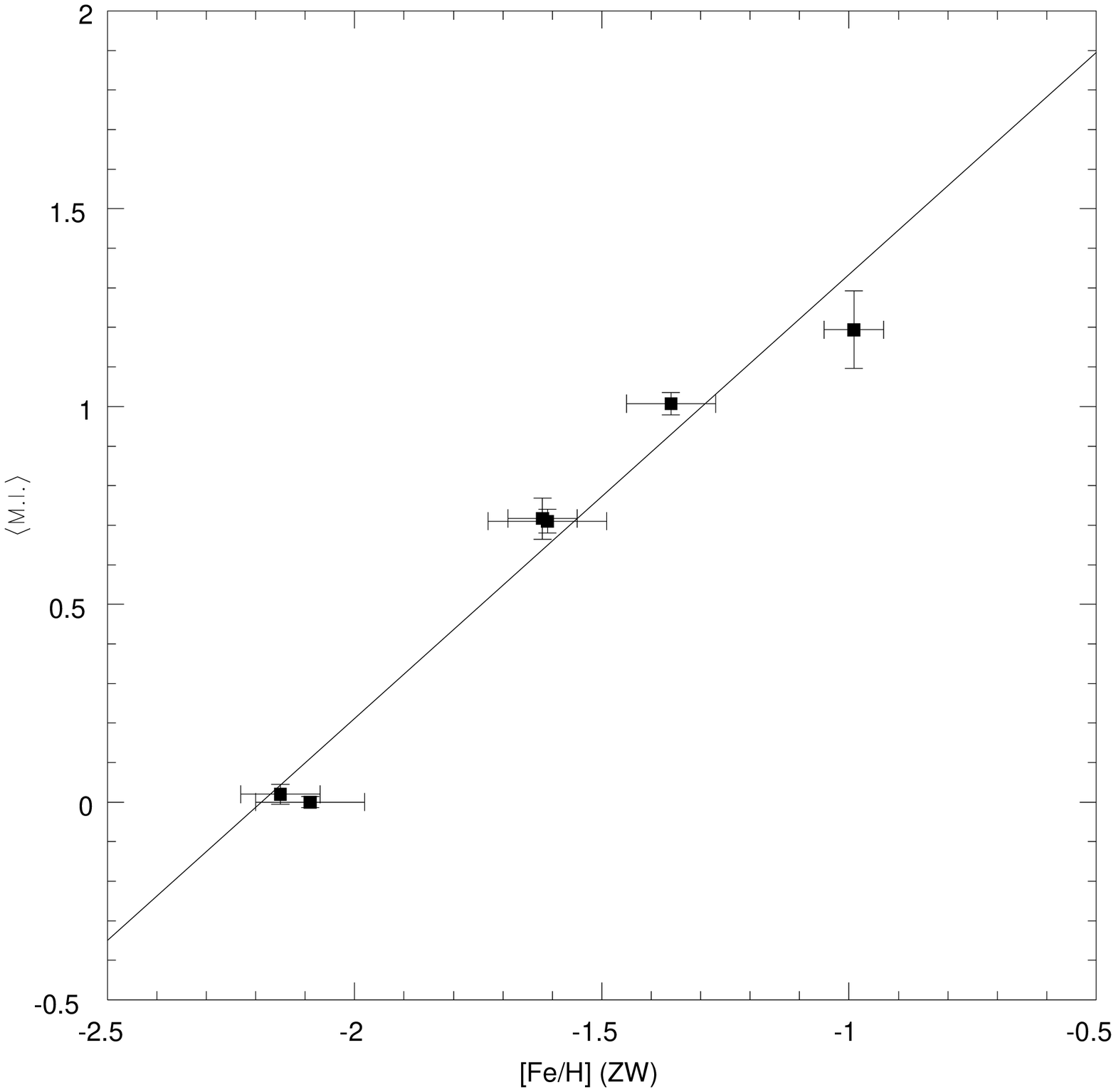}
\includegraphics[width=8.8cm]{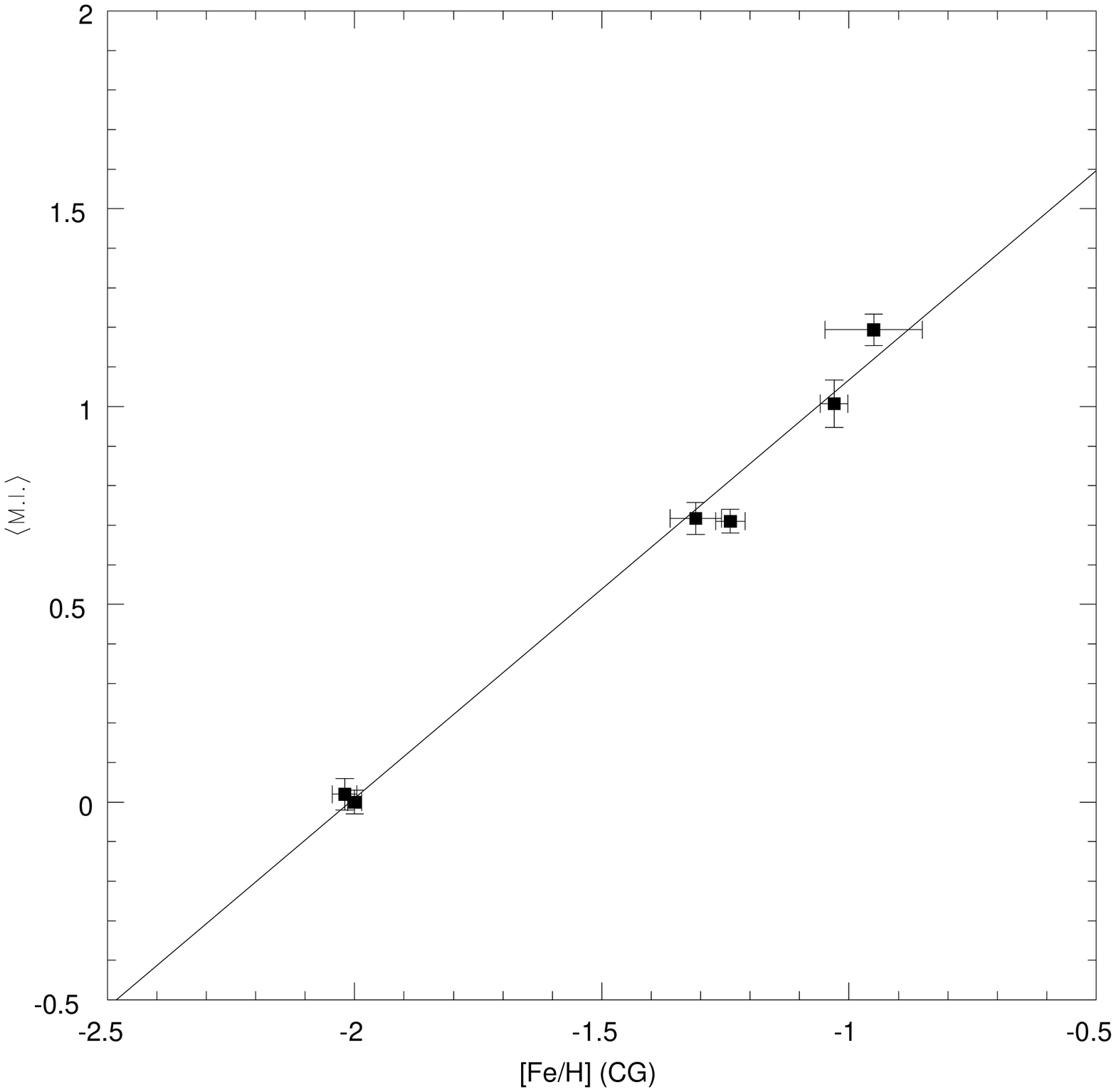}
{Figure 8. Calibration of the metallicity index (M.I.) on ZW84 
(left panel), and CG97 (right panel) metallicity scales.
}
\label{f:cal02}
\end{figure*}
\noindent
and are shown in Fig.~8
for the ZW84 and CG97 metallicity scales, respectively.

Individual metal abundances for the RR Lyrae stars in the calibrating clusters
analyzed in the present paper were derived from the above calibration equations.
They 
are listed in 
Tables~\ref{t:M15}, ~\ref{t:M2}, ~\ref{t:NGC6171} respectively, while 
the mean metallicities derived from the averages 
of these individual values and their respective dispersions are given 
in 
Table~\ref{t:calibration}.

The last column in each of Tables~\ref{t:M15},~\ref{t:M2}, and ~\ref{t:NGC6171}
 gives individual radial velocities measured from the 
spectra of the stars observed in the calibrating clusters. Averages of these
values are listed in 
Table~\ref{t:calibration}.
We note that these average radial velocities differ somewhat from the literature
values, particularly for NGC\,6171, suggesting the presence of 
systematic offsets possibly caused by offcentering of the cluster variables
in the slit. These systematic differences are not found in the case of the
Sculptor variables, for which the average radial velocity we measured
is perfectly consistent with the literature values (see Section 5). 
This suggests that we are simply seeing an effect of the small samples 
in the Galactic clusters, while the one in Sculptor is large enough to
average away the offcentering effects.

\end{document}